\documentclass[preprint,showpacs,showkeys,preprintnumbers,amsmath,amssymb]{revtex4}
\usepackage[usenames,dvipsnames]{color}
\usepackage{graphicx}
\usepackage{dcolumn}
\usepackage{bm}
\usepackage{color}
\usepackage{bm}
\usepackage{amsmath}
\usepackage{graphicx}
\usepackage{dsfont}

\def\|{|\!|}

\newcommand{\ba}{\begin{array}}\newcommand{\ea}{\end{array}}
\newcommand{\be}{\begin{equation}}\newcommand{\ee}{\end{equation}}
\newcommand{\bea}{\begin{eqnarray}}\newcommand{\eea}{\end{eqnarray}}
\newcommand{\brr}{\begin{array}}\newcommand{\err}{\end{array}}
\newcommand{\bit}{\begin{itemize}}\newcommand{\eit}{\end{itemize}}
\newcommand{\ben}{\begin{enumerate}}\newcommand{\een}{\end{enumerate}}

\renewcommand{\baselinestretch}{1.1}

\definecolor{darkred}{rgb}{.8,0,0}

\definecolor{darkblue}{rgb}{0,0,.7}

\def\boldsymbol#1{{\bf #1}}%

\def\beq{\begin{equation}}
\def\eeq{\end{equation}}

\begin{document}

\title{On the Role of Information Theoretic Uncertainty Relations in Quantum Theory\\}

\author{Petr~Jizba}
 \email{p.jizba@fjfi.cvut.cz}
\affiliation{FNSPE, Czech Technical University in Prague,
B\v{r}ehov\'{a} 7, 115 19 Praha 1, Czech Republic\\ {\rm and} \\
ITP, Freie Universit\"{a}t Berlin, Arnimallee 14 D-14195 Berlin,
Germany\\}

\author{Jacob A. Dunningham}
\email{J.Dunningham@sussex.ac.uk}
\affiliation{Department of Physics and Astronomy, University of Sussex, Falmer, Brighton, BN1 9QH UK\\}

\author{Jaewoo Joo}
\email{J.Joo@leeds.ac.uk}
\affiliation{School of Physics and Astronomy, University of Leeds, Leeds LS2 9JT, UK\\}

\begin{abstract}
Uncertainty relations based on information theory for both discrete
and continuous distribution functions are briefly reviewed. We
extend these results to account for (differential) R\'{e}nyi
entropy and its related entropy power. This allows us to find a new class
of information-theoretic uncertainty relations (ITURs). The potency of such
uncertainty relations in quantum mechanics is illustrated with a
simple two-energy-level model where they outperform both the usual
Robertson--Schr\"{o}dinger uncertainty relation and Kraus--Maassen
Shannon entropy based uncertainty relation. In the continuous case
the ensuing entropy power uncertainty relations are discussed in the context of
heavy tailed wave functions and Schr\"odinger cat states.
Again, improvement over both the Robertson--Schr\"{o}dinger
uncertainty principle and Shannon ITUR is demonstrated in these cases. Further salient
issues such as the proof of a generalized entropy power inequality and
a geometric picture of information-theoretic uncertainty relations are
also discussed.
\end{abstract}


\maketitle

\section{Introduction}

Quantum-mechanical uncertainty relations place fundamental limits on the
accuracy with which one is able to measure the values of different physical
quantities. This has profound implications not only on the microscopic but
also on the macroscopic level of physical systems.
The archetypal uncertainty relation formulated by Heisenberg in 1927 describes a trade-off
between the error of a measurement to know the value of one observable and the
disturbance caused on another complementary observable so that
their product should be no less than a limit set by $\hbar$.
Since Heisenberg's intuitive, physically motivated deduction of the
error-disturbance uncertainty relations~\cite{Heis,Heis2}, a number of methodologies trying to improve
or supersede this result have been proposed. In fact, over the years it have became steadily clear that
the intuitiveness of Heisenberg's version cannot substitute mathematical rigor and it came  as
no surprise that the violation of the Heisenberg's original relation was
recently reported a number of experimental groups, e.g., most recently by the Vienna group
in neutron spin measurements~\cite{Ozawa:12}. At present it is
Ozawa's universally valid error-disturbance relation~\cite{Ozawa:03,Ozawa:05} that represents a viable
alternative to Heisenberg's error-disturbance relation.

Yet, already at the end of
1920s Kennard and independently Robertson and Schr\"{o}dinger reformulated
the original  Heisenberg (single experiment, simultaneous measurement, error-disturbance)
uncertainty principle in terms of a statistical ensemble of identically
prepared experiments~\cite{Kenn1,Robertson1929,Schroedinger1}. Among other things, this
provided a rigorous meaning to Heisenberg's imprecisions (``Ungenauigkeiten")
$\delta x$ and $\delta p$ as standard
deviations in position and momenta, respectively, and entirely avoided the
troublesome concept of \emph{simultaneous} measurement.
The Robertson--Schr\"{o}dinger approach has proven to be sufficiently versatile
to accommodate other complementary observables apart from $x$ and $p$,
such as components of angular momenta, or energy and time.
Because in the above cases the variance is taken as a \textquotedblleft
measure of uncertainty", expressions of this type are also known as
variance-based uncertainty relations (VUR). Since Robertson and
Schr\"{o}dinger's papers, a multitude of VURs has been devised; examples include the
Fourier-type uncertainty relations of Bohr and Wigner~\cite{Dirac1958,Donoho1989},
the fractional Fourier-type uncertainty relations of Mustard~\cite{Mustard:91}, mixed-states
uncertainty relations~\cite{Dodonov1980}, the angle-angular momentum uncertainty
relation of L\'{e}vy-Leblond~\cite{Levy-Leblond1976} and Carruthers and
Nietto~\cite{Carruthers1968}, the time-energy uncertainty relation of
Mandelstam and Tamm~\cite{Mandelstam1945}, Luisell's amplitude-phase
uncertainty relation~\cite{Luisell:73},  and Synge's three-observable
uncertainty relations~\cite{Synge1971}.

Many authors~\cite{Hirschman1957, Bialynicky-Birula1975,
Deutsch1983, Maassen1988, Uffink1990, Montgomery2002} have, however,
remarked that even VURs have many limitations. In fact, the essence of  a VUR
is to put an upper bound to the degree of concentration of two (or
more) probability distributions, or, equivalently impose a lower
bound to the associated uncertainties. While the variance is often a
good measure of the concentration of a given distribution, there are
many situations where this is not the case. For instance, variance
as a measure of concentration is a dubious concept in the case when
a distribution contains more than one peak. Besides, variance
diverges in many distributions even though such distributions are
sharply peaked. Notorious examples of the latter are provided by
heavy-tail distributions such as L\'{e}vy~\cite{levy,feller}, Weibull~\cite{feller} or
Cauchy--Lorentz
distributions~\cite{feller, C-L distribution}. For instance, in the theory of Bright--Wigner
shapes it has been known for a long time~\cite{Lyons1989} that the
Cauchy--Lorentz distribution can be freely concentrated into an
arbitrarily small region by changing its scale parameter, while its
standard deviation remains very large or even infinite.

Another troublesome feature of VURs appears in the case of finite-dimensional
Hilbert spaces, such as the Hilbert space of spin or angular momentum.
The uncertainty product can attain zero minimum even when one
of the distributions is not absolutely localized, i.e., even when the value of
one of the observables is not precisely known~\cite{Deutsch1983}. In such a
case the uncertainty is just characterized by the lower bound of the
uncertainty product (i.e., by zero) and thus it only says that this product is
greater than zero for some states and equal to zero for others. This is,
however, true also in classical physics.

The previous examples suggest that it might be desirable to quantify
the inherent quantum unpredictability in a different, more expedient way.
A distinct class of such non-variance-based uncertainty relations are the uncertainty
relations based on information theory. In these the uncertainty is
quantified in terms of various information measures --- entropies,
which often provide more stringent bound on concentrations of
the probability distributions. The purpose of the present paper
is to give a brief account of the existing information-theoretic
uncertainty relations (ITUR) and present some new results based on
R\'{e}nyi entropy. We also wish to promote the
notion of R\'{e}nyi entropy (RE) which is not yet sufficiently
well known in the physics community.

Our paper is organized in the following way: In Section~\ref{I}, we provide
some information-theoretic background on the R\'{e}nyi entropy (RE).
In particular, we stress distinctions between the RE for discrete probabilities and RE for continuous
probability density functions (PDF) --- the so-called differential RE.
In  Section~\ref{II} we briefly review the concept of {\em entropy power} both for Shannon and R\'{e}nyi entropy.
We also prove the generalized entropy power inequality.
%
%
With the help of the Riesz--Thorin inequality we derive in Section~\ref{Sec3}
the RE-based ITUR for discrete distributions. In addition, we also propose a geometric illustration of
the latter  in terms of the {\em condition number} and
{\em distance to singularity}.
In Section~\ref{SEc4} we employ the Beckner--Babenko inequality
to derive a continuous variant of the RE-based ITUR. The result is phrased both
in the language of REs and generalized entropy powers. In particular, the
latter allows us to establish a logical link with the Robertson--Schr\"{o}dinger VUR.
The advantage of ITURs over the usual
VUR approach is  illustrated in Section~\ref{SEc5}.
In two associated subsections we first examine the r\^{o}le of a discrete generalized ITUR
on a simple two-level quantum system.
In the second subsection the continuous ITUR is considered for quantum-mechanical systems with heavy-tailed
distributions and Schr\"{o}dinger cat states.
An improvement of the R\'{e}nyi ITUR over both the Robertson--Schr\"{o}dinger VUR and
Shannon ITUR is demonstrated in all the cases discussed. Finally in Section~\ref{SEc6} we make some concluding remarks and
propose some generalizations. For the reader's convenience we relegate to Appendix~A some of the detailed
mathematical steps needed in Sections~\ref{II.a} and \ref{SEc4}.

\section{Brief introduction to R\'{e}nyi entropy~\label{I}}

The basic notion that will be repeatedly used in the following
sections is the notion of R\'{e}nyi entropy. For this reason we
begin here with a brief review of some of its fundamental properties.

REs constitute a one-parameter family of information entropies
labeled by R\'{e}nyi's parameter $\alpha\in\mathbb{R}^{+}$ and
fulfill additivity with respect to the composition of
statistically independent systems. The special case with $\alpha=1$
corresponds to the familiar Shannon entropy. It can be shown that
R\'{e}nyi entropies belong to the class of mixing homomorphic
functions~\cite{Lesche1982} and that they are analytic for
$\alpha$'s which lie in $I\cup IV$ quadrants of the complex
plane~\cite{Jizba2004}. In order to address the uncertainty
relations issue it is important to distinguish two situations.

\subsection{Discrete probability distribution case~\label{I.a}}

Let $\mathcal{X}=\{x_{1},\ldots,x_{n}\}$ be a random variable
admitting $n$ different events (be it outcomes of some experiment or
microstates of a given macrosystem), and let
$\mathcal{P}=\{p_{1},\ldots,p_{n}\}$ be the corresponding
probability distribution. Information theory then ensures that the
most general information measures (i.e. entropy) compatible with
the additivity of independent events are those of
R\'{e}nyi~\cite{Renyi1970}:
\begin{equation}
\mathcal{I}_{\alpha}(\mathcal{P})=\frac{1}{(1-\alpha)}\,\log_{2}\left(
\sum_{k=1}^{n}p_{k}^{\alpha}\right)  \,. \label{ren11}
\end{equation}
Form (\ref{ren11}) is valid even in the limiting case when
$n\rightarrow\infty$. If, however, $n$ is finite then R\'{e}nyi
entropies are bounded both from below and from above:
$\log_{2}(p_{k})_{\max}\leq \mathcal{I}_{\alpha}\leq\log_{2}n.$ In
addition, REs are monotonically decreasing functions in $\alpha$,
so $\mathcal{I}_{\alpha_{1}}<\mathcal{I}_{\alpha_{2}}$ if and
only if $\alpha_{1}>\alpha_{2}$. One can reconstruct the entire
underlying probability distribution knowing all R\'{e}nyi
distributions via the Widder--Stiltjes inverse formula~\cite{Jizba2004}.
In this case the leading order contribution comes from
$\mathcal{I}_{1}(\mathcal{P})$, i.e. from Shannon's entropy. Some
elementary properties of $\mathcal{I}_{\alpha}$ are as follows:

\begin{enumerate}
\item RE is symmetric: ${\mathcal{I}}_{\alpha}(p_{1},\ldots,p_{n})
={\mathcal{I}}_{\alpha}(p_{k(1)},\ldots,p_{k(n)})$\thinspace\ .

\item RE is nonnegative: ${\mathcal{I}}_{\alpha}({\mathcal{P}})\geq
0$\thinspace\ .

\item RE is decisive: ${\mathcal{I}}_{\alpha}(0,1)={\mathcal{I}}_{\alpha
}(1,0)$ \thinspace\ .

\item For ${\alpha}\leq1$ RE is concave; for $\alpha>1$
RE in neither convex nor concave.

\item RE is bounded, continuous and monotonic in $\alpha$\thinspace\ .

\item RE is analytic in ${\alpha}\in{\mathbb{C}}_{I\cup III}$ $\Rightarrow$
for ${\alpha}=1$ it equals to Shannon's entropy, i.e.
$\lim_{{\alpha
}\rightarrow1}{\mathcal{I}}_{\alpha}={\mathcal{H}}$.\newline
\end{enumerate}

\noindent Among a myriad of information measures REs distinguish
themselves by having a firm operational characterization in terms of
block coding and hypotheses testing. R\'{e}nyi's parameter $\alpha$
is then directly related to so-called $\beta$-cutoff
rates~\cite{Csisz'ar1995}. RE is used in coding theory~\cite{Campbell1965,baer-2006},
cryptography~\cite{Nielsen2000,Cachin1997,Benett:95}, finance~\cite{Jizba:12,Jizba:14} and in theory of
statistical inference~\cite{Renyi1970}. In physics one often uses $\mathcal{I}_{\alpha}(\mathcal{P})$
in the framework of quantum information theory~\cite{Benett:95,Hayashi:11,Adesso:12}.


\subsection{Continuous probability distribution case~\label{I.b}}

Let $M$ be a measurable set on which is defined a continuous probability density function (PDF)
$\mathcal{F}({\boldsymbol{x}})$. We will assume that the support (or
outcome space) is a smooth but not necessarily compact manifold. By
covering the support with the mesh $M^{(l)}$ of $d$--dimensional
(disjoint) cubes $M_{k}^{(l)}$ $(k=1,\ldots,n)$ of size $l^{d}$ we
may define the integrated probability in $k$--th cube as
\begin{equation}
p_{nk}=\mathcal{F}({\boldsymbol{x}}_{i})l^{d}\,,\;\;\;{\boldsymbol{x}}_{i}\in
M_{k}^{(l)}\,.
\end{equation}
This defines the mesh probability distribution
$\mathcal{P}_{n}=\{p_{n1},\ldots,p_{nn}\}$. Infinite precision of
measurements (i.e., when $l\rightarrow0$) often brings infinite
information. As the most \textquotedblleft junk" information comes
from the uniform distribution $\mathcal{E}_{n}$, it is more sensible
to consider the relative information entropy rather than absolute
one. In references~\cite{Jizba2004, Renyi1970} it was shown that in
the limit $n\rightarrow\infty$ (i.e., $l\rightarrow0$) it is
possible to define a finite information measure compatible with
information theory axioms. This \emph{renormalized} R\'{e}nyi
entropy, often known as {\em differential RE entropy}, reads
\begin{equation}
\tilde{\mathcal{I}}_{\alpha}(\mathcal{F})\equiv\lim_{n\rightarrow\infty
}(\mathcal{I}_{\alpha}(\mathcal{P}_{n})-\mathcal{I}_{\alpha}(\mathcal{E}%
_{n}))\ =\ \frac{1}{(1-\alpha)}\,\log_{2}\left(  \frac{\int_{M}%
d{{\boldsymbol{x}}}\,\mathcal{F}^{\alpha}({\boldsymbol{x}})}{\int
_{M}d{{\boldsymbol{x}}}\,1/V^{\alpha}}\right)  \,. \label{ren55}%
\end{equation}
Here $V$ is the volume of $M$. Equation (\ref{ren55}) can be viewed
as a generalization of the Kullback--Leibler relative
entropy~\cite{Kullback51}.
When $M$ is compact it is possible to
introduce a simpler alternative prescription as
\begin{align}
\mathcal{I}_{\alpha}(\mathcal{F})  &  \equiv\lim_{n\rightarrow\infty
}(\mathcal{I}_{\alpha}(\mathcal{P}_{n})-\mathcal{I}_{\alpha}(\mathcal{E}
_{n})|_{V=1})\ =\ \lim_{n\rightarrow\infty}(\mathcal{I}_{\alpha}
(\mathcal{P}_{n})+D\log_{2}l)\nonumber\\
&  =\frac{1}{(1-\alpha)}\log_{2}\left(  \int_{M}d{{\boldsymbol{x}}
}\,\mathcal{F}^{\alpha}({\boldsymbol{x}})\right)  \,. \label{ren6}
\end{align}
In both previous cases
$D$ represents the Euclidean dimension of the support. R\'{e}nyi
entropies (\ref{ren55}) and (\ref{ren6}) are defined if (and only if) the
corresponding integral
$\int_{M}d{{\boldsymbol{x}}}\,\mathcal{F}^{\alpha}({\boldsymbol{x}})$
exists. Equations (\ref{ren55}) and (\ref{ren6}) indicate that
the asymptotic expansion for $\mathcal{I}_{\alpha}(\mathcal{P}_{n})$
has the form:
\[
\mathcal{I}_{\alpha}(\mathcal{P}_{n})\ =\
-D\log_{2}l+\mathcal{I}_{\alpha
}(\mathcal{F})+o(1)\ =\ -D\log_{2}l+\tilde{\mathcal{I}}_{\alpha}%
(\mathcal{F})+\log_{2}V_{n}+\mathcal{O}(1)\,.
\]
\noindent Here $V_{n}$ is the covering volume and the symbol $\mathcal{O}(1)$
is the residual error which tends to $0$ for $l\rightarrow0$. In
contrast to the discrete case, R\'{e}nyi entropies
$\mathcal{I}_{\alpha}(\mathcal{F})$ are not generally positive.
In particular, a distribution which is more confined than a unit volume has less RE
than the corresponding entropy of a uniform distribution over a unit volume
and hence  yields a negative $\mathcal{I}_{\alpha}(\mathcal{F})$. A paradigmatic example
of this type of behavior is the $\delta$-function PDF in which case
$\mathcal{I}_{\alpha} = -\log_2 \delta(0) = -\infty$, for all $\alpha$.
Information measures $\tilde{\mathcal{I}}_{\alpha }(\mathcal{F})$
and $\mathcal{I}_{\alpha}(\mathcal{F})$ are often applied in theory
of statistical inference~\cite{Arimitsu2000, Arimitsu2001,
Arimitsu2002, Arimitsu2002a} and in chaotic dynamical
systems~\cite{Halsey1986,Jensen1985,Tomita1988,Hentschel1983}.

\section{Entropy power and entropy power inequalities~\label{II}}

The mathematical underpinning for most uncertainty relations used in
quantum mechanics lies in inequality theory. For example, the
wave-packet uncertainty relations are derived from the Plancherel
inequality, and the celebrated Robertson--Schr\"{o}dinger's VUR is based on
the Cauchy--Schwarz inequality (and ensuing Parseval
equality)~\cite{Robertson1929}. Similarly, Fourier-type uncertainty
relations are based on the Hausdorff--Young
inequality~\cite{Andersson1993}, etc.

In information theory the key related inequalities are a) Young's inequality that
implies the entropy power inequalities, b) the Riesz--Thorin inequality that
determines the generalized entropic
uncertainty relations and c) the Cram\'{e}r--Rao and logarithmic Sobolev inequalities
that imply Fisher's information uncertainty principle. In this section we
will briefly review the concept of the {\em entropy power}
and the ensuing {\em entropy power inequality}. Both concepts were developed by Shannon in his
seminal 1948 paper in order to bound the capacity of non-Gaussian additive noise
channels~\cite{Shannon48}. The connection with quantum mechanics was established by Stam~\cite{Stam:59}, Lieb~\cite{Lieb1976}
and others who used the entropy power inequality to prove standard VUR.

In the second part of this section we show how the entropy power can be extended into the RE setting.
With the help of Young's inequality we find the corresponding generalized entropy power inequality.
Related applications to quantum mechanics will be postponed to Section~\ref{SEc5.b}.


%

\subsection{Entropy power inequality --- Shannon entropy case\label{II.a}}

Suppose that $\mathcal{X}$ is a random vector in $\mathbb{R}^{D}$ with
the PDF $\mathcal{F}$. The differential (or
continuous) entropy $\mathcal{H}(\mathcal{X})$ of $\mathcal{X}$ is defined
as
\begin{equation}
\mathcal{H}(\mathcal{X})\ = \ \mathcal{I}_{1}(\mathcal{F})\ = \ -\int_{
\mathbb{R}^{D}}\mathcal{F}({\boldsymbol{x}})\log_{2}\mathcal{F}%
({\boldsymbol{x}})\ d{\boldsymbol{x}}\,. \label{IIA1}%
\end{equation}
The discrete version of (\ref{IIA1}) is nothing but the Shannon
entropy~\cite{Shannon48}, and in such a case it represents an
average number of binary questions that are needed to reveal the
value of ${\mathcal{X}}$. Actually, (\ref{IIA1}) is not a proper
entropy but rather information gain~\cite{Jizba2004, Renyi1970} as can be seen directly from
(\ref{ren6}) when the limit $\alpha  \rightarrow 1$ is taken.
We
shall return to this point in Section~5. The \emph{entropy power}
$N(\mathcal{X})$ of ${\mathcal{X}}$ is the unique number
such that~\cite{Shannon48, Costa1985}
\begin{eqnarray}
{\mathcal{H}} \left( {\mathcal{X}} \right) \
= \ \mathcal{H}\left({\mathcal{X}}_{G}\right)\, ,
\label{3.1.0b}
\end{eqnarray}
where ${\mathcal{X}}_{G}$ is a Gaussian random vector with zero mean and
variance equal to $N(\mathcal{X})$, i.e., ${\mathcal{X}}_{G} \sim {\mathcal{N}}({\boldsymbol{0}}, N(\mathcal{X}) \mathds{1}_{D\times D})$. Eq.(\ref{3.1.0b}) can be equivalently rewritten in the form
\begin{eqnarray}
{\mathcal{H}} \left( {\mathcal{X}} \right) \
= \ \mathcal{H}\left(\sqrt{N(\mathcal{X})}\cdot {\mathcal{Z}}_{G}\right) ,
\label{3.1.0bv}
\end{eqnarray}
with ${\mathcal{Z}}_{G}$ representing a Gaussian random vector with the zero mean and unit covariance matrix.
The solution of both
(\ref{3.1.0b}) and (\ref{3.1.0bv}) is then
\begin{eqnarray}
N(\mathcal{X})\ = \ \frac{2^{ \frac{2}{D}\ \!
\mathcal{H}(\mathcal{X}) }}{2\pi e} \,
.
\label{A.8.a}
\end{eqnarray}
Let $\mathcal{X}_{1}$ and $\mathcal{X}_{2}$ be two independent continuous
vector valued random variables of finite variance. In the case when the Shannon differential entropy is measured in
{\em nats} (and not bits) we get for the entropy power
\begin{eqnarray}
N(\mathcal{X})\ = \ \frac{1}{2\pi e}\exp\left( \frac{2}{D}\ \!
\mathcal{H}(\mathcal{X})\right)  \,
.
\label{A.8.b}
\end{eqnarray}
The differential
entropy (\ref{A.8.a}) (as well as (\ref{A.8.b}))  satisfies the so-called  {\em entropy power inequality}
\begin{eqnarray}
N(\mathcal{X}_{1} + \mathcal{X}_{2})\ \geq \ N(\mathcal{X}_{1})
\ + \ N(\mathcal{X}_{1})\, , \label{3.1.0}
\end{eqnarray}
where the equality holds iff $\mathcal{X}_{1}$ and $\mathcal{X}_{2}$ are
multivariate normal random variables with proportional covariance
matrices~\cite{Shannon48}. In general, inequality (\ref{3.1.0}) does not hold when
$\mathcal{X}_{1}$ and $\mathcal{X}_{2}$ are discrete random variables and the differential
entropy is replaced with the discrete entropy. Shannon originally used this inequality to
obtain a lower bound for the capacity of non-Gaussian additive noise channels.
Since Shannon's pioneering paper several proofs of the entropy power inequality have
become available~\cite{CovThom,Verdu2006,Stam:59,Blachman:65}.

\subsection{Entropy power inequality --- R\'{e}nyi entropy case\label{II.ab}}

In the following we will show how it is possible to extend the entropy power
concept to REs. To this end we first define R\'{e}nyi entropy power (for simplicity we use nats
as units of information).

\newtheorem{definition}{Definion}[section]
\begin{definition} Let $p>1$ and let ${\mathcal{X}}$ be a random vector
in ${\mathbb{R}^{D}}$ with probability density
${\mathcal{F}} \in \ell^{p}({\mathbb{R}^{D}})$. The $p$-th R\'{e}nyi entropy power of
${\mathcal{X}}$ is defined as
\begin{eqnarray}
N_p(\mathcal{X})\ = \ \frac{1}{2\pi} p^{-p'/p} |\!|
{\mathcal{F}}|\!|_p^{-2p'/D} \ = \ \frac{1}{2\pi} p^{-p'/p}
\exp\left(\frac{2}{D} \ \!{\mathcal{I}}_p({\mathcal{F}})\right),
\label{3.1.0e}
\end{eqnarray}
where $p'$ is the H\"{o}lder conjugate of $p$.
\end{definition}

The above form of $N_p(\mathcal{X})$ was probably firstly stated
by Gardner~\cite{Gardner02} who, however, did not develop the analogy with
$N(\mathcal{X})$ any further. Plausibility of
$N_p(\mathcal{X})$ as  the entropy power comes from the
following important properties:

\newtheorem{theorem}{Theorem}[section]
\begin{theorem}
The $p$-th R\'{e}nyi entropy power $N_p(\mathcal{X})$ is a
unique solution of the equation
\begin{eqnarray}
{\mathcal{I}_{p}} \left( {\mathcal{X}} \right)
\ = \ \mathcal{I}_{p}\left(\sqrt{N_p(\mathcal{X})}\cdot {\mathcal{Z}}_{G}\right)\,
.\label{3.1.0k}
\end{eqnarray}
With ${\mathcal{Z}}_{G}$ representing a Gaussian random vector with zero mean and unit covariance matrix.
In addition, in the limit $p \rightarrow 1_+$ one has
$N_p(\mathcal{X}) \rightarrow N(\mathcal{X})$.

Let $\mathcal{X}_{1}$ and $\mathcal{X}_{2}$ be two independent
continuous random vectors in $\mathbb{R}^{D}$ with probability
densities ${\mathcal{F}}^{(1)} \in \ell^{q}({\mathbb{R}^{D}})$ and
${\mathcal{F}}^{(2)} \in \ell^{p}({\mathbb{R}^{D}})$, respectively.
Suppose further that $\lambda \in (0,1)$ and $r>1$, and let
\begin{eqnarray*}
q = \frac{r}{(1-\lambda) + \lambda r}\, , \;\;\;\; p =
\frac{r}{\lambda + (1-\lambda) r}\, .
\end{eqnarray*}
Then the following inequality holds:
\begin{eqnarray}
{N}_{r}(\mathcal{X}_{1}+\mathcal{X}_{2}) \ \geq \ \left(\frac{{N}_{q}
(\mathcal{X}_{1})}{1-\lambda} \right)^{1-\lambda} \left(\frac{{N}_{p}
(\mathcal{X}_{2})}{\lambda} \right)^{\lambda}. \label{3.1.0a}
\end{eqnarray}
Additionally, in the limits $r,p,q \rightarrow 1_+$  the inequality (\ref{3.1.0a})
reduces to the Shannon entropy power inequality (\ref{3.1.0})  and
${N}_{1}(\mathcal{X}) = {N}(\mathcal{X})$.\label{th1}
\end{theorem}
\vspace{3mm}

\noindent {\em Proof of Theorem~\ref{th1}.}

That $N_p(\mathcal{X})$ from Definition~III.1 is the only solution of (\ref{3.1.0k}) follows from the scaling property of
${\mathcal{I}_{p}}$, namely
\begin{eqnarray}
{\mathcal{I}_{p}}(a \mathcal{X}) \ = \   {\mathcal{I}_{p}}(\mathcal{X}) \ + \ D\log_2 |a|\, ,
\label{14.bb}
\end{eqnarray}
where $a \in \mathbb{R}$. The above scaling relation follows directly from the definition of ${\mathcal{I}_{p}}$ and from a change of  variable argument. We can further use the simple fact that
\begin{eqnarray}
{\mathcal{I}_{p}}({\mathcal{Z}}_{G}) \ = \  \frac{D}{2} \log_2(2\pi p^{p'/p})\, ,
\label{15.bb}
\end{eqnarray}
to see that (\ref{3.1.0k}) leads to the equation
\begin{eqnarray}
{\mathcal{I}_{p}}(\mathcal{X}) \ = \ \frac{D}{2} \log_2\left(2\pi p^{p'/p} N_p(\mathcal{X}) \right).
\end{eqnarray}
This yields
\begin{eqnarray}
N_p(\mathcal{X}) \ = \ \frac{1}{2\pi}\ \!p^{-p'/p} \ \! \  \! 2^{\frac{2}{D}\ \! {\mathcal{I}_{p}}(\mathcal{X})}\, ,
\end{eqnarray}
which, for ${\mathcal{I}_{p}}$ measured in {\em nats}, coincides with (\ref{3.1.0e}).

To prove the inequality (\ref{3.1.0a}) we first realize that $p$,
$q$ and $r$ represent H\"{o}lder's triple, i.e.
\begin{eqnarray}
\frac{1}{q} + \frac{1}{p} = 1 + \frac{1}{r}\, . \label{3.1.0c}
\end{eqnarray}
This allows us to use Young's inequality (q.v. Appendix~A), which for the case
at hand reads
\begin{eqnarray}
|\!|{\mathcal{F}}^{(1)}\ast {\mathcal{F}}^{(2)}|\!|_r \leq C^D
|\!|{\mathcal{F}}^{(1)}|\!|_q|\!|{\mathcal{F}}^{(2)}|\!|_p\, ,
\label{3.1.0d}
\end{eqnarray}
where $C$ is a constant defined in Appendix~A.
The left-hand-side of (\ref{3.1.0d}) can be explicitly written as
\begin{eqnarray}
|\!|{\mathcal{F}}^{(1)}\ast {\mathcal{F}}^{(2)}|\!|_r =
\left[\int_{{\mathbb{R}^{D}}} d{{\boldsymbol{x}}} \left(\int_
{{\mathbb{R}^{D}}}d{{\boldsymbol{y}}} \ \!
{\mathcal{F}}^{(1)}({\boldsymbol{x}} - {\boldsymbol{y}})
{\mathcal{F}}^{(2)}({\boldsymbol{y}}) \right)^{\!r}\right]^{1/r} .
\label{3.1.0f}
\end{eqnarray}
The probability ${\mathcal{F}}^{(1)}({\boldsymbol{x}} -
{\boldsymbol{y}}) {\mathcal{F}}^{(2)}({\boldsymbol{y}})$ is nothing
but the joint probability that $\mathcal{X}_1 = {\boldsymbol{x}} -
{\boldsymbol{y}} $ and $\mathcal{X}_2 = {\boldsymbol{y}}$. The quantity
inside $(\ldots)$  thus  represents the density function for the sum
of two random variables $\mathcal{X}_1 + \mathcal{X}_2 =
{\boldsymbol{x}}$. With the help of (\ref{3.1.0e}) we can rewrite
(\ref{3.1.0f}) as
\begin{eqnarray}
|\!|{\mathcal{F}}^{(1)}\ast {\mathcal{F}}^{(2)}|\!|_r = [2\pi
{N}_{r}(\mathcal{X}_{1}+\mathcal{X}_{2}) ]^{-D/2r'}
r^{-D/2r}. \label{3.1.0g}
\end{eqnarray}
On the other hand, the right-hand-side of (\ref{3.1.0d}) is
\begin{eqnarray}
|\!|{\mathcal{F}}^{(1)}|\!|_q|\!|{\mathcal{F}}^{(2)}|\!|_p =
 [2\pi
{N}_{q}(\mathcal{X}_{1})]^{-D/2q'}[2\pi
{N}_{p}(\mathcal{X}_{2})]^{-D/2p'} q^{-D/2q} p^{-D/2p}\, .
\label{3.1.0h}
\end{eqnarray}
Plugging (\ref{3.1.0g}) and (\ref{3.1.0h}) into the Young inequality
(\ref{3.1.0d}) we obtain
\begin{eqnarray}
{N}_{r}(\mathcal{X}_{1}+\mathcal{X}_{2}) &\geq&
|r'|^{-1}|q'|^{-r'/q'}|p'|^{-r'/p'}
[{N}_{q}(\mathcal{X}_{1})]^{r'/q'}
[{N}_{p}(\mathcal{X}_{2})]^{r'/p'}\nonumber \\[2mm]
&=& \left(\frac{{N}_{q}(\mathcal{X}_{1})}{1-\lambda
}\right)^{1-\lambda}
\left(\frac{{N}_{p}(\mathcal{X}_{2})}{\lambda}
\right)^{\lambda}.
\end{eqnarray}
This completes the proof of the inequality (\ref{3.1.0a}).

It remains to show that in the limits $r,p,q \rightarrow 1_+$ we
regain the Shannon entropy power inequality. Firstly, the above
limits directly give the inequality
\begin{eqnarray}
{N}(\mathcal{X}_{1}+\mathcal{X}_{2}) &\geq&
\left(\frac{{N}(\mathcal{X}_{1})}{1-\lambda
}\right)^{1-\lambda}
\left(\frac{{N}(\mathcal{X}_{2})}{\lambda}
\right)^{\lambda}, \label{3.1.0i}
\end{eqnarray}
which holds without restrictions on $\lambda \in (0,1)$. The
best estimate (the highest lower bound) is obtained for $\lambda$
that extremizes the right-hand-side. Assuming that the
right-hand-side is for fixed $\mathcal{X}_{1}$ and $\mathcal{X}_{2}$
a smooth function of $\lambda$, we can take its derivative with respect to
$\lambda$. This equals zero when
\begin{eqnarray}
{N}(\mathcal{X}_{1}) =
\left(\frac{1-\lambda}{\lambda}\right){N}(\mathcal{X}_{2})
\;\;\;\;\; \Leftrightarrow \;\;\;\;\;\lambda =
\frac{{N}(\mathcal{X}_{2})}{{N}(\mathcal{X}_{1})
+ {N}(\mathcal{X}_{2})} \, . \label{3.1.0j}
\end{eqnarray}
Positivity of ${N}(\ldots)$ then ensures that $\lambda$,
which extremizes the right-hand-side of (\ref{3.1.0i}), belongs to
the interval $(0,1)$. In addition, the extremum is actually a maximum
because the second derivative is $-[{N}(\mathcal{X}_{1}) +
{N}(\mathcal{X}_{2})]^3/{N}(\mathcal{X}_{1}){N}(\mathcal{X}_{2})$
which is clearly negative. By inserting (\ref{3.1.0j}) into (\ref{3.1.0i}) we regain the Shannon
entropy power inequality.

To prove that $N(\mathcal{X})$ is a limiting case of
$N_p(\mathcal{X})$ for $p \rightarrow 1_+$,
we just realize that $p^{-p'/p} \rightarrow 1/e$ and
$|\!| {\mathcal{F}}|\!|_p^{-2p'/D} \rightarrow \exp\left(\frac{2}{D}\ \!
{\mathcal{I}}_1({\mathcal{F}})\right)$. Thus indeed
in the $p \rightarrow 1_+$ limit we regain the original Shannon
entropy power $N(\mathcal{X})$ as well as the usual entropy power inequality (\ref{3.1.0}).  $~~~~\square$\\[0mm]

In passing we may observe that from the definition (\ref{3.1.0e})
and Eqs.~(\ref{14.bb})-(\ref{15.bb}) it follows that
$N_p({\sigma}\mathcal{Z}_G) = \sigma^2$, i.e. the power entropy
coincides for Gaussian processes with the variance $\sigma^2$. In
case when $\mathcal{Z}_G$ represents a random Gaussian vector of
zero mean and covariance matrix $\bf{K}$, then $N_p(\mathcal{Z}_G) =
|{\bf{K}}|^{1/D}$. Note that these statements are $p$-independent
and hence valid also for the original Shannon entropy power.

\section{Information Theoretic Uncertainty Relations and R\'{e}nyi entropy -
discrete distributions\label{Sec3}}

\subsection{The Riesz--Thorin Inequality~\label{III.a}}

To prove the information uncertainty relation based on RE we need to prove a particular variant of the Riesz--Thorin
inequality~\cite{Riesz1926, Thorin1939, Hardy1959} upon which our
considerations will be based. For this purpose we first state the Riesz
convexity theorem.

\begin{theorem}
[Riesz convexity theorem]Let ${\mathcal{L}}$ be a linear operator (i.e.,
$({\mathcal{L}} {\boldsymbol{x}})_{j} = \sum_{i} a_{ij} x_{j}$) and $|\!|
{\boldsymbol{y}}|\!|_{p} = \left(  \sum_{i} | y_{i} |^{p} \right)  ^{1/p} $.
Let, in addition, $M_{\alpha\beta}$ be the least number ``$k$" satisfying
\begin{align*}
|\!| {\mathcal{L}}{\boldsymbol{x}}|\!|_{1/(1-\beta)} \, \leq\, k |\!|
{\boldsymbol{x}}|\!|_{1/\alpha}.
\end{align*}
Then $\log(M_{\alpha\beta})$ is convex in triangle $0 \leq\alpha;\beta
\leq1$, $\, \alpha+ \beta\geq1$.
\end{theorem}

\begin{figure}
[ptb]
\begin{center}
\includegraphics[
height=1.913in, width=2.1975in] {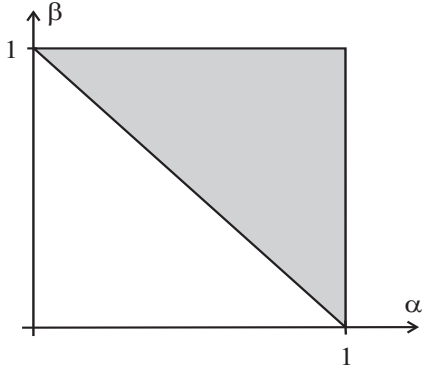} \caption{Riesz convexity
triangle. Riesz's inequality in Theorem~1 holds only when $\alpha$
and $\beta$ belong to the shaded region.}%
\label{ConcFig}
\vspace{0.2cm} \hrule
\end{center}
\end{figure}

The convexity triangle is depicted in Figure~\ref{ConcFig}. Detailed
exposition of the proof can be found for example in~\cite{Hardy1959}.

\newtheorem{corollary}{Corollary}[section]
\begin{corollary}
Let ($\alpha_{1},\beta_{1}$) and ($\alpha_{2},\beta_{2}$) be two points in the
above convex triangle. If we define
\[
\alpha=\alpha_{1}s+\alpha_{2}(1-s)\,,\,\,\,\,\beta=\beta_{1}s+\beta
_{2}(1-s)\,;\;\;\;\;\;\;\;s\in\lbrack0,1]\,,
\]
then clearly
\begin{eqnarray*}
\log(M_{\alpha \beta}) \ \leq \ s \log(M_{\alpha_1 \beta_1}) + (1-s)
\log(M_{\alpha_2 \beta_2 })\, ,
\end{eqnarray*}
or equivalently
\[
M_{\alpha\beta}\ \leq\ M_{\alpha_{1}\beta_{1}}^{s}M_{\alpha_{2}\beta_{2}%
}^{(1-s)}\,.
\]
\end{corollary}

\begin{theorem}
[Riesz--Thorin inequality]Suppose that $({\mathcal{L}}{\boldsymbol{x}}%
)_{j}=\sum_{i}a_{ji}x_{i}$ and that
\[
\sum_{j}|({\mathcal{L}}{\boldsymbol{x}})_{j}|^{2}\ \leq\ \sum_{j}|x_{j}%
|^{2}\, .
\]
Then for $p\in\lbrack1,2]$ and $c\equiv\max_{i,j}|a_{ij}|$
\[
|\!|{\mathcal{L}}{\boldsymbol{x}}|\!|_{p^{\prime}}\,\leq\,c^{(2-p)/p}%
|\!|{\boldsymbol{x}}|\!|_{p}\,=\,c^{1/p}c^{-1/p^{\prime}}|\!|{\boldsymbol{x}%
}|\!|_{p}\;\;\;\Leftrightarrow\;\;\;c^{1/p^{\prime}}|\!|{\mathcal{L}%
}{\boldsymbol{x}}|\!|_{p^{\prime}}\,\leq\,c^{1/p}|\!|{\boldsymbol{x}}%
|\!|_{p}\,,
\]
holds. Here $p$ and $p^{\prime}$ are H\"{o}lder conjugates, i.e., ${1}/
{p}+{1}/{p^{\prime}}=1$. \label{th2}
\end{theorem}
\vspace{3mm}

\noindent {\em Proof of Theorem~\ref{th2}.} We shall use the
notation $\alpha=1/p$, $\beta=1/q$ (and the H\"{o}lder conjugates
$p^{\prime}=p/(p-1)$, $q^{\prime}=q/(q-1)$). Consider the line from
$(\alpha_{1},\beta_{1})=(1/2,1/2)$ to $(\alpha_{2},\beta_{2})=(1,1)$
in the $(\alpha,\beta)$ plane. This line lies entirely in the
triangle of concavity (see Figure \ref{ConcFig}). Let us now define
\begin{align*}
\alpha & =\alpha_{1}s+\alpha_{2}(1-s)\\
& =s/2+(1-s)\\
& =-s/2+1\, ,
\end{align*}
implying $s=2(1-\alpha)$, and define
\begin{align*}
\beta & =\beta_{1}s+\beta_{2}(1-s)\\
& = - s/2 + 1\, ,
\end{align*}
implying $\beta=\alpha.$ Hence
\begin{equation}
M_{\alpha,\alpha}\,\leq\,M_{\alpha_{1}\beta_{1}}^{s}M_{\alpha_{2}\beta_{2}%
}^{(1-s)}\,=\,M_{1/2,1/2}^{2(1-\alpha)}\ M_{1,1}^{2\alpha-1}\,.\label{proof2}%
\end{equation}
Note particularly that because $s\in\lbrack0,1]$ then
$\alpha\in\lbrack1/2,1]$ and $p\in\lbrack1,2]$. To estimate the right
hand side of (\ref{proof2}) we first realize that
$M_{1/2,1/2}\leq1$. This results from the very assumption of the
theorem, namely that
\[
\left\Vert \mathcal{L}{\boldsymbol{x}}\right\Vert _{2}^2=\sum_{j}|(\mathcal{L}%
{\boldsymbol{x}})_{j}|^{2}\,\leq\,\sum_{j}|x_{j}|^{2}=\left\Vert
{\boldsymbol{x}}\right\Vert _{2}^2\,.
\]
Hence, $M_{1/2,1/2}\leq k=1$. To find the estimate for $M_{11}$ we realize that
it represents the smallest  $k$ in the relation
\[
|\!|\mathcal{L}{\boldsymbol{x}}|\!|_{\infty}\,\leq\,k|\!|{\boldsymbol{x}}%
|\!|_{1}\,.
\]
Thus
\begin{eqnarray}
M_{11}\ =\ \max_{{\boldsymbol{x}}\neq 0}\frac{|\!|\mathcal{L}{\boldsymbol{x}
}|\!|_{\infty}}{|\!|{\boldsymbol{x}}|\!|_{1}}\ =\ \max_{{\boldsymbol{x}}\neq 0}\frac
{\max_{j}|(\mathcal{L}{\boldsymbol{x}})_{j}|}{\sum_{i}|x_{i}|}
\leq
\ \max_{i,j}|a_{ij}|\ \equiv \ c\,.
\label{A.22}
\end{eqnarray}
So finally we can write that
\begin{eqnarray*}
\mbox{\hspace{3
cm}}M_{\alpha,\alpha}\ =\ M_{1/p,(1-1/p^{\prime})} \
\leq\ c^{2\alpha-1} \ =\ c^{(2-p)/p}\ =\
c^{1/p}c^{-1/p^{\prime}}\,.\mbox{\hspace{2.5cm}} \Box
\end{eqnarray*}
%

\subsection{Generalized ITUR~\label{III.b}}

\begin{figure}
[ptb]
\begin{center}
\includegraphics[
height=2.3601in, width=2.6662in ] {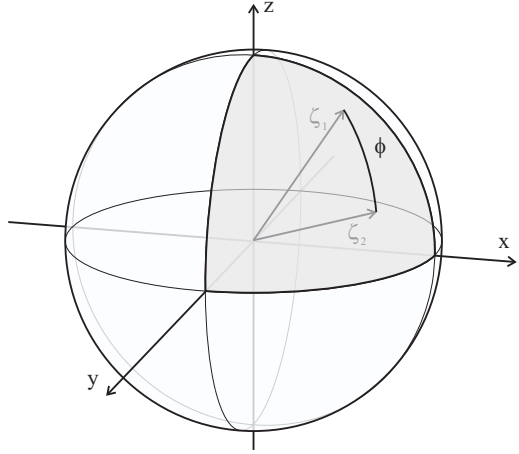} \caption{A statistical
system can be represented by points $\xi$ on a positive orthant
$(S^{n-1})^{+}$ of the unit sphere $S^{n-1}$ in a real Hilbert space
$H$. The depicted example corresponds to $n=3$. } \label{fig1}
\vspace{0.2cm} \hrule
\end{center}
\end{figure}

To establish the connection with RE let us assume that $\mathcal{X}$ is a
discrete random variable with $n$ different values, $\mathbb{P}_{\!n}$ is the
probability space affiliated with $\mathcal{X}$ and $\mathcal{P}%
=\{p_{1},\ldots,p_{n}\}$ is a sample probability distribution from
$\mathbb{P}_{\!n}$. Normally the geometry of $\mathbb{P}_{\!n}$ is identified
with the geometry of a simplex. For our purpose it is more interesting to
embed $\mathbb{P}_{\!n}$ in a sphere. Because $\mathcal{P}$ is non--negative
and summable to unity, it follows that the square--root likelihood $\xi
_{i}=\sqrt{p_{i}}$ exists for all $i=1,\ldots,n$, and it satisfies the
normalization condition
\[
\sum_{i=1}^{n}(\xi_{i})^{2}=1\,.
\]
Hence ${\boldsymbol{\xi}}$ can be regarded as a unit vector in the Hilbert
space $\mathcal{H}=\mathbb{R}^{n}$. Then the inner product
\begin{eqnarray}
\cos\phi=\sum_{i=1}^{n}\xi_{i}^{(1)}\xi_{i}^{(2)}=1-\frac{1}{2}\sum_{i=1}%
^{n}\left(  \xi_{i}^{(1)}-\xi_{i}^{(2)}\right)  ^{2}\, , \label{4.5}
\end{eqnarray}
defines the angle $\phi$ that can be interpreted as a distance
between two probability distributions. More precisely, if
$\mathcal{S}^{n-1}$ is the unit sphere in the $n$-dimensional
Hilbert space, then $\phi$ is the spherical (or geodesic) distance
between the points on $\mathcal{S}^{n-1}$ determined by
${\boldsymbol{\xi}}^{(1)}$ and ${\boldsymbol{\xi}}^{(2)}$. Clearly,
the maximal possible distance, corresponding to orthogonal
distributions, is given by $\phi=\pi/2$. This follows from the fact
that ${\boldsymbol{\xi}}^{(1)}$ and ${\boldsymbol{\xi}}^{(2)}$ are
non--negative, and hence they are located only on the positive
orthant of $\mathcal{S}^{n-1}$ (see Figure~\ref{fig1}).
The geodesic distance
$\phi$ is called the Bhattacharyya distance. The representation
of probability distributions as points on a sphere also has an
interesting relation to Bayesian statistics. If we use a uniform
distribution on the sphere as the prior distribution then the prior
distribution on probability vectors in $\mathbb{P}_{\!n}$ is exactly
the celebrated Jeffrey's prior that has found new justification via
the minimum description length approach to statistics~\cite{Grunwald2007}.

Now, let $\mathcal{P}^{(1)}$ and $\mathcal{P}^{(2)}$ denote a pair
of probability distributions and ${\boldsymbol{\xi}}^{(1)}$ and
${\boldsymbol{\xi }}^{(2)}$ the corresponding elements in Hilbert
space. Because ${\boldsymbol{\xi}}^{(1)}$ and
${\boldsymbol{\xi}}^{(2)}$ are non-negative, they are located only
on the positive orthant of $\mathcal{S}^{n-1}$. The transformation
$\mathcal{L}{{\boldsymbol{\xi}}}^{(1)}={\boldsymbol{\xi}}^{2}$ then
corresponds to a rotation with $a_{ij}\in SO(n)$.

To proceed, we set $p^{\prime}=2(1+t)$ and $p=2(1+r)$ (remembering that
$1/p+1/p^{\prime}=1$). Then the Riesz--Thorin inequality reads (with ${\xi}^{(1)}_i
\leftrightarrow x_i$)
\begin{equation}
\left(  \sum_{i}(\xi_{i}^{(2)})^{p^{\prime}}\right)  ^{1/p^{\prime}}
\ \leq\ c^{(2-p)/p}\left(  \sum_{i}(\xi_{i}^{(1)})^{p}\right)
^{1/p},\label{3.3.2}
\end{equation}
which is equivalent to
\[
\left(  \sum_{j}(p_{j}^{(2)})^{(1+t)}\right)  ^{1/2(1+t)}\left(  \sum
_{k}(p_{k}^{(1)})^{(1+r)}\right)  ^{-1/2(1+r)}\ \leq\ c^{-r/(1+r)}.
\]
We raise both sides to the power $2\left( 1+t\right)  /t$ and get
\begin{equation}
\left(  \sum_{j}(p_{j}^{(2)})^{(1+t)}\right)  ^{1/t}\left(  \sum_{k}
(p_{k}^{(1)})^{(1+r)}\right)  ^{-(1+t)/t(1+r)}\ \leq\
c^{-2r(1+t)/t(1+r)}. \label{Ineq}
\end{equation}
The parameters are limited due to the condition $p\in\lbrack1,2]$
and $1/p+1/p^{\prime}=1$ implying
\begin{equation}
t=-r/(2r+1). \label{Iden}
\end{equation}
This implies that $r\in\lbrack-1/2,0]$ and $t\in\lbrack0,\infty)$.
Combining (\ref{Ineq}) and (\ref{Iden}) we get
\begin{equation}
\left(  \sum_{j}(p_{j}^{(2)})^{(1+t)}\right)  ^{1/t}\left( \sum_{k}
(p_{k}^{(1)})^{(1+r)}\right)  ^{1/r}\leq\ c^{2}. \label{Prae}
\end{equation}
By applying the negative binary logarithm on both sides of
(\ref{Prae}) we get the following theorem.

\begin{theorem}
\label{hoved}Suppose that $({\mathcal{L}}{\boldsymbol{x}})_{j}=\sum_{i}%
a_{ij}x_{j} \equiv (\mathbb{A}  {\boldsymbol{x} })_j$ and that
\[
\sum_{j}|({\mathcal{L}}{\boldsymbol{x}})_{j}|^{2}\ \leq\ \sum_{j}|x_{i}%
|^{2}\,,\;\;\;\ \text{for all}\;\;x_{i}\,.
\]
Define $c\equiv\max_{i,j}|a_{ij}|.$ If $r\in\lbrack-1/2,0]$ and
$t=-r/(2r+1)$ and the probability distributions $\mathcal{P}^{(1)}$
and $\mathcal{P}^{(2)}$ are related by
$\mathcal{L}{{\boldsymbol{\xi}}}^{(1)}={\boldsymbol{\xi}}^{\left(
2\right)  }$ where $\xi_{i}=\sqrt{p_{i}}$, then
\begin{eqnarray}
\mathcal{I}_{1+t}(\mathcal{P}^{(2)})+\mathcal{I}_{1+r}(\mathcal{P}^{(1)})\
\geq\ -2\log_{2}c\, . \label{3.3.1}
\end{eqnarray}

\end{theorem}

Two immediate comments are in order. Firstly, one can extend
the domain of validity of both $r$ and $t$ by noticing that
$\mathcal{P}^{(1)}$ and $\mathcal{P}^{(2)}$ are interchangeable in
the above derivation without altering~\cite{footnote1} the actual value of $c$. This has the
consequence that one may phrase both resultant inequalities as a
single inequality where both $r$ and $t$ belong to the interval
$[-1/2, \infty)$ with $t = - r/(2r + 1)$. Secondly, because the
information measure $\mathcal{I}_{\alpha}(\mathcal{P})$ is always
non-negative, the inequality (\ref{3.3.1}) can represent a genuine
uncertainty relation only when $c < 1$.
Note that for $\mathbb{A} \in SO(n)$ or $SU(n)$ (i.e. for most physically relevant situations) one always has
that $c\leq 1$. This is because for such $\mathbb{A}$'s
\begin{eqnarray}
c \ = \ \max_{i,j}|a_{ik}| \ = \ |\! |\mathbb{A}|\!|_{\rm{max}} \ \leq \ |\! |\mathbb{A}|\!|_2 \ = \ \sqrt{\lambda_{\rm{max}}(\mathbb{A}^{\dag}\mathbb{A})} \ = \ 1\, .
\end{eqnarray}
The last identity results from the fact that all of eigenvalues of
 $\mathbb{A} \in SO(n)$ or $SU(n)$ have absolute value $1$.

It needs to be stressed that in the particular case when $r=0$
(and thus also $t=0$) we get
\begin{equation}
\mathcal{H}(\mathcal{P}^{(2)})\ + \ \mathcal{H}(\mathcal{P}^{(1)})\
\geq \ -2\log _{2}c\, .  \label{kendt}
\end{equation}
This Shannon entropy based uncertainty relation  was originally
found by Kraus~\cite{Kraus87} and Maassen~\cite{Maassen1988}. A
weaker version of this ITUR was also earlier proposed by
Deutsch~\cite{Deutsch1983}.

The reader can see that ITUR  (\ref{3.3.1}) which is based on
RE provides a natural extension of the Shannon ITUR
(\ref{kendt}). In Section~\ref{SEc5} we shall see that there are quantum
mechanical systems where R\'{e}nyi's ITUR improves both on
Robertson--Schr\"{o}dinger's VUR and Shannon's ITUR.

\subsection{Geometric interpretation of inequality (\ref{3.3.1})
~\label{III.c}}

Let us close this section by providing a useful
geometric understanding of the inequality (\ref{3.3.1}).
To this end we invoke two concepts known from  error analysis. These are, the
{\em condition number} and {\em distance to singularity} (see, e.g., Refs.~\cite{Golub:96,Steward:90}).

The {\em condition number}
 $\kappa_{\alpha,\beta}(\mathbb{A})$ of the non-singular matrix $\mathbb{A}$
is defined as
\begin{eqnarray}
\kappa_{\alpha,\beta}(\mathbb{A}) \ = \ |\!| \mathbb{A} |\!|_{\alpha,\beta} |\!| \mathbb{A}^{-1} |\!|_{\beta, \alpha}\, ,
\end{eqnarray}
where, the corresponding (mixed) matrix-valued norm $|\!| \mathbb{A} |\!|_{\alpha,\beta}$ is defined as
\begin{eqnarray}
|\!| \mathbb{A} |\!|_{\alpha,\beta} \ = \ \max_{{\boldsymbol{x}}\neq 0} \frac{|\!| \mathbb{A} {\boldsymbol{x}} |\!|_{\beta} }{|\!|{\boldsymbol{x}}|\!|_{\alpha}}\, .
\end{eqnarray}
So, in particular $M_{11} = c$ from (\ref{A.22}) is nothing but $|\!| \mathbb{A} |\!|_{1,\infty}$.
Note also that $|\!| \mathbb{A} |\!|_{\alpha,\alpha} = |\!| \mathbb{A} |\!|_{\alpha}$, which is the usual $\alpha$-matrix norm.
Justification for calling $\kappa_{\alpha,\beta}$ a condition number comes from the following
theorem:

\begin{theorem}
Let $\mathbb{A} {\boldsymbol{x}} = {\boldsymbol{y}}$ be a linear equation and let there be an error
(or uncertainty) $\delta {\boldsymbol{y}}$  in representing the vector ${\boldsymbol{y}}$, and let $\hat{\boldsymbol{x}} = {\boldsymbol{x}} + \delta {\boldsymbol{x}}$ solve the new  error-hindered equation
$\mathbb{A} \hat{\boldsymbol{x}} = {\boldsymbol{y}} + \delta {\boldsymbol{y}}$. The relative disturbance in ${\boldsymbol{x}}$ in relation to $\delta {\boldsymbol{y}}$ fulfills
\begin{eqnarray}
\frac{|\!|\delta{\boldsymbol{x}}|\!|_{\alpha}}{|\!|{\boldsymbol{x}}|\!|_{\alpha}} \ \leq \ \kappa_{\alpha,\beta}(\mathbb{A}) \frac{|\!|\delta{\boldsymbol{y}}|\!|_{\beta}}{|\!|{\boldsymbol{y}}|\!|_{\beta}}\, .
\label{33a}
\end{eqnarray}
\label{th4}
\end{theorem}

\noindent {\em Proof of Theorem~\ref{th4}.} The proof is rather simple. Using the fact that $\mathbb{A} {\boldsymbol{x}} = {\boldsymbol{y}}$ and
$\mathbb{A} \hat{\boldsymbol{x}} = {\boldsymbol{y}} + \delta {\boldsymbol{y}}$ we obtain $\delta {\boldsymbol{x}} = \mathbb{A}^{-1} \delta {\boldsymbol{y}}$. Taking $\alpha$-norm on both sides we can write
\begin{eqnarray}
|\!|\delta {\boldsymbol{x}}|\!|_{\alpha}  \ =  \
|\!|\mathbb{A}^{-1} \delta {\boldsymbol{y}}|\!|_{\alpha}\ \leq \ |\!|\mathbb{A}^{-1} |\!|_{\beta,\alpha}
|\!|\delta {\boldsymbol{y}}|\!|_{\beta}\, .
\label{34a}
\end{eqnarray}
On the other hand, the $\beta$-norm of $\mathbb{A} {\boldsymbol{x}} = {\boldsymbol{y}}$ yields
\begin{eqnarray}
|\!| {\boldsymbol{y}}|\!|_{\beta} \ = \  |\!| \mathbb{A} {\boldsymbol{x}} |\!|_{\beta} \ \leq \ |\!| \mathbb{A}  |\!|_{\alpha, \beta}  |\!|{\boldsymbol{x}} |\!|_{\alpha}\;\;\;\;\;\Leftrightarrow \;\;\;\;\;
\frac{1}{|\!|{\boldsymbol{x}} |\!|_{\alpha} } \ \leq \ \frac{|\!| \mathbb{A}  |\!|_{\alpha, \beta} }{|\!| {\boldsymbol{y}}|\!|_{\beta}}\, .
\label{35a}
\end{eqnarray}
Combining (\ref{34a}) with (\ref{35a}) we obtain (\ref{33a}).  $\square$\\[0mm]

From the previous theorem we see that $\kappa_{\alpha,\beta}(\mathbb{A})$ quantifies a {\em stability} of the linear equation $\mathbb{A} {\boldsymbol{x}} = {\boldsymbol{y}}$, or better the extent to which the relative error (uncertainty) in ${\boldsymbol{y}}$ influences the relative error in ${\boldsymbol{x}}$. A system described by $\mathbb{A}$ and ${\boldsymbol{y}}$ is stable if $\kappa_{\alpha,\beta}(\mathbb{A})$ is not too large (ideally close to one). It is worth of stressing that  $\kappa_{\alpha,\beta}(\mathbb{A}) \geq 1$.
The latter results from the fact that
\begin{eqnarray}
|\!| {\boldsymbol{x}}|\!|_{\beta} \ = \ |\!| \mathbb{A} \mathbb{A}^{-1}{\boldsymbol{x}}|\!|_{\beta}
\ \leq \ |\!| \mathbb{A} \mathbb{A}^{-1}|\!|_{\beta,\beta} |\!| {\boldsymbol{x}}|\!|_{\beta}
\ \leq \ |\!| \mathbb{A} |\!|_{\alpha,\beta} |\!| \mathbb{A}^{-1} |\!|_{\beta, \alpha}|\!| {\boldsymbol{x}}|\!|_{\beta}\, .
\end{eqnarray}
In the last step we have used the submultiplicative property of mixed matrix norms.

The second concept --- the {\em distance to singularity} for a matrix $\mathbb{A}$, is defined  as
\begin{eqnarray}
\mbox{dist}_{\alpha,\beta}(\mathbb{A})\ \equiv \ \mbox{min}\left\{|\!|\Delta \mathbb{A}
|\!|_{\alpha,\beta}; \,\, \mathbb{A} + \Delta \mathbb{A} \,\, {\mbox{singular}}  \right\}.
\end{eqnarray}
In this connection an important theorem states that the {\em relative} distance to
singularity is the reciprocal of the condition number.

\begin{theorem}
For a non-singular matrix $\mathbb{A}$, one has
\begin{eqnarray}
\frac{{\mbox{\rm{dist}}}_{\alpha,\beta}(\mathbb{A})}{|\!| \mathbb{A} |\!|_{\alpha,\beta}} \ = \ \kappa_{\alpha,\beta}(\mathbb{A})^{-1}\, .
\label{4.38a}
\end{eqnarray}
\label{th5}
\end{theorem}

\noindent {\em Proof of Theorem~\ref{th5}.} If $\mathbb{A} + \Delta \mathbb{A}$ is singular then there is a
vector ${\boldsymbol{x}} \neq 0$, such that $(\mathbb{A} + \Delta \mathbb{A}){\boldsymbol{x}} = 0$. Because
$\mathbb{A}$ is non-singular, the latter is equivalent to ${\boldsymbol{x}} =  - \mathbb{A}^{-1}\Delta \mathbb{A} {\boldsymbol{x}}$. By taking the $\alpha$-norm we have
\begin{eqnarray}
|\!|{\boldsymbol{x}}|\!|_{\alpha} \ = \ |\!|\mathbb{A}^{-1}\Delta \mathbb{A} {\boldsymbol{x}}|\!|_{\alpha}
\ \leq \ |\!|\mathbb{A}^{-1}|\!|_{\beta, \alpha }|\!|\Delta \mathbb{A} {\boldsymbol{x}}|\!|_{\beta}
\ \leq \ |\!|\mathbb{A}^{-1}|\!|_{\beta, \alpha }|\!|\Delta \mathbb{A}|\!|_{\alpha,\beta} |\!|{\boldsymbol{x}}|\!|_{\alpha}\, ,
\end{eqnarray}
which is equivalent to
\begin{eqnarray}
\frac{|\!|\Delta \mathbb{A}|\!|_{\alpha,\beta} }{|\!|\mathbb{A}|\!|_{\alpha, \beta }} \ \geq \ \kappa_{\alpha,\beta}(\mathbb{A})^{-1}\, .
\label{4.40a}
\end{eqnarray}
To show that $\kappa_{\alpha,\beta}(\mathbb{A})^{-1}$ is a true minimum of the left-hand side of
(\ref{4.40a}) and not mere lower bound we must show that there exists such a suitable perturbation $\Delta \mathbb{A}$ which saturates the inequality.
Corresponding $|\!|\Delta \mathbb{A}|\!|_{\alpha,\beta}$ will then clearly represent
$\mbox{dist}_{\alpha,\beta}(\mathbb{A})$. Consider ${\boldsymbol{y}}$ such that $|\!|{\boldsymbol{y}}|\!|_{\beta} = 1$ and $|\!|\mathbb{A}^{-1}{\boldsymbol{y}}|\!|_{\alpha} = |\!|\mathbb{A}^{-1}|\!|_{\beta, \alpha }$, and write ${\boldsymbol{z}} = \mathbb{A}^{-1} {\boldsymbol{y}}$.
Define further a vector $\hat{{\boldsymbol{z}}}$ such that
\begin{eqnarray}
\max_{|\!|{\boldsymbol{\zeta}}|\!|_{\alpha} = 1}\frac{|\hat{\boldsymbol{z}}^*\cdot {\bf{\zeta}} |}{|\!|{\boldsymbol{z}}|\!|_{\alpha}} \ = \ \frac{\hat{{\boldsymbol{z}}}^*\cdot {\boldsymbol{z}}}{|\!|{\boldsymbol{z}}|\!|_{\alpha}} \ = \ 1\, .
\end{eqnarray}
We now introduce the matrix $\mathbb{B}_{i,j} = - {\boldsymbol{y}}_i \hat{\boldsymbol{z}}^*_j$, which implies $\mathbb{B}{\boldsymbol{z}}/|\!|{\boldsymbol{z}}|\!|_{\alpha} = -{\boldsymbol{y}}$. Note that $\mathbb{B}$ thus defined fulfills
\begin{eqnarray}
|\!|\mathbb{B}|\!|_{\alpha,\beta} \ = \ \max_{|\!|{\boldsymbol{\zeta}}|\!|_{\alpha} = 1} \frac{|\!|{\boldsymbol{y}}(\hat{\boldsymbol{z}}^*\cdot {\boldsymbol{\zeta}})|\!|_{\beta}}{|\!|{\boldsymbol{z}}|\!|_{\alpha}} \ = \ |\!|{\boldsymbol{y}}|\!|_{\beta}
\max_{|\!|{\boldsymbol{\zeta}}|\!|_{\alpha} = 1} \frac{|\hat{\boldsymbol{z}}^*\cdot {\bf{\zeta}} |}{|\!|{\boldsymbol{z}}|\!|_{\alpha}} \ = \ 1\, .
\end{eqnarray}
Let us set $\Delta \mathbb{A} = \mathbb{B}/|\!|{\boldsymbol{z}}|\!|_{\alpha}$. This directly implies that
\begin{eqnarray}
(\mathbb{A} + \Delta \mathbb{A})\mathbb{A}^{-1}{\boldsymbol{y}} \ = \ {\boldsymbol{y}} + \frac{\mathbb{B}{\boldsymbol{z}}}{|\!|{\boldsymbol{z}}|\!|_{\alpha}} \ = \ 0\, .
\end{eqnarray}
So the matrix $\mathbb{A} + \Delta \mathbb{A}$ is singular with $\mathbb{A}^{-1}{\boldsymbol{y}}$ being the null vector. Finally note that
\begin{eqnarray*}
\mbox{\hspace{3cm} }\frac{|\!|\Delta \mathbb{A}|\!|_{\alpha,\beta} }{|\!|\mathbb{A}|\!|_{\alpha, \beta }} \ = \ \frac{|\!| \mathbb{B}|\!|_{\alpha,\beta} }{|\!|{\boldsymbol{z}}|\!|_{\alpha}|\!|\mathbb{A}|\!|_{\alpha, \beta }} \ = \
\frac{1}{|\!|\mathbb{A}^{-1}{\boldsymbol{y}}|\!|_{\alpha}|\!|\mathbb{A}|\!|_{\alpha, \beta }} \ = \ \kappa_{\alpha,\beta}(\mathbb{A})^{-1}\, . \mbox{\hspace{2.2cm} }\square
\end{eqnarray*}
\\
The connection with the ITUR (\ref{3.3.1}) is established when we observe that the smallest value of $c$ is (see, Eqs.(\ref{A.22}) and (\ref{4.38a}))
\begin{eqnarray}
c \ = \  |\!|\mathbb{A}|\!|_{1, \infty} \ = \ \mbox{dist}_{1,\infty}(\mathbb{A}) \ \! \kappa_{1,\infty}(\mathbb{A})\, .
\end{eqnarray}
Since $c\leq1$, this shows that the ITUR (\ref{3.3.1}) restricts the probability distributions more the smaller the distance to
singularity and/or the lower the stability of the transformation matrix
$\mathbb{A}$  is.
In practical terms this means that the
rotation/transformation within the positive orthant
introduces higher ignorance or uncertainty in the ITUR
the more singular the rotation/transformation matrices are.
\section{Information Theoretic Uncertainty Relations
and R\'{e}nyi entropy - continuous distributions~\label{SEc4}}

Before considering  quantum-mechanical implications of
R\'{e}nyi's ITUR (\ref{3.3.1}), we will briefly touch upon the
continuous-probability analogue of (\ref{3.3.1}). This issue is
conceptually far more delicate than the discrete one namely because
it is difficult to find norms for the correspondent (integro-)differential
operators ${\mathcal{L}}$. This in particular does not allow one
to calculate explicitly the optimal bounds in many relevant cases.
Fortunately, there is one very important class of situations, where one can
proceed with relative ease. This is the situation when the linear
transform is represented by a continuous Fourier transform, in which
case the Riesz-Thorin inequality is taken over by the Beckner-Babebko
inequality~\cite{Beckner1975,Babenko1962}.

\begin{theorem} [Beckner--Babebko's theorem]
Let
\[
f^{(2)}({{\boldsymbol{x}}}) \equiv
\hat{f}^{(1)}({{\boldsymbol{x}}})= \int_{\mathbb{R}^{D}}e^{2\pi
i{{\boldsymbol{x}} }.{{\boldsymbol{y}}}}\
f^{(1)}({{\boldsymbol{y}}})\ d{{\boldsymbol{y}}}\,,
\]
then for $p \in [1,2]$ we have
\begin{eqnarray}
|\!|\hat{f}|\!|_{p^{\prime}}\ \leq\
\frac{|p^{D/2}|^{1/p}}{|(p^{\prime} )^{D/2}|^{1/p^{\prime}}}\
|\!|f|\!|_{p}\, , \label{4.1}
\end{eqnarray}
or, equivalently
\begin{eqnarray*}
\;|(p^{\prime})^{D/2}|^{1/p^{\prime}}|\!|f^{(2)}|\!|_{p^{\prime}} \
\leq\ |p^{D/2}|^{1/p}|\!|f^{(1)}|\!|_{p}\, .
\end{eqnarray*}
Here, again, $p$ and $p^{\prime}$ are the usual H\"{o}lder
conjugates. For any  $F \in
\ell^{p}({\mathbb{R}^{D}})$ the $p$-norm $|\!|F|\!|_{p}$ is defined as
\[
|\!|F|\!|_{p} = \left(\int_{\mathbb{R}^{D}}
|F({\boldsymbol{y}})|^{p} \ d{{\boldsymbol{y}}}\right)^{1/p}.
\]
Due to symmetry of the Fourier transform the reverse inequality also holds:
\begin{eqnarray}
|\!|{f}|\!|_{p^{\prime}}\ \leq\
\frac{|p^{D/2}|^{1/p}}{|(p^{\prime} )^{D/2}|^{1/p^{\prime}}}\
|\!|\hat{f}|\!|_{p}\, . \label{4.1b}
\end{eqnarray}
\end{theorem}

The proof of this theorem can be found in the Appendix.
Lieb~\cite{Lieb1990} proved that the inequality (\ref{4.1}) is
saturated only for Gaussian functions. In the case of discrete
Fourier transforms the corresponding inequality is known as the
(classical) Hausdorff--Young inequality~\cite{Hardy1959, Andersson1993}.

Analogous manipulations that have brought us from equation
(\ref{3.3.2}) to equation (\ref{Prae}) will allow us to cast
(\ref{4.1}) in the form
\begin{eqnarray}
\left(\int_{\mathbb{R}^{D}} [
{\mathcal{F}}^{(2)}({\boldsymbol{y}})]^{(1+t)} \
d{{\boldsymbol{y}}}\right)^{1/t}\left(\int_{\mathbb{R}^{D}}
[{\mathcal{F}}^{(1)}({\boldsymbol{y}})]^{(1+r)} \
d{{\boldsymbol{y}}} \right)^{1/r} \leq\ [2(1+t)]^D \left|t/r
\right|^{D/2r}, \label{4.2}
\end{eqnarray}
where we have defined the square-root density likelihood as
$|f({\boldsymbol{y}})| = \sqrt{{\mathcal{F}}({\boldsymbol{y}})}$.

\begin{figure}
[t]
\begin{center}
\includegraphics[height=2.3801in, width=3.4662in ] {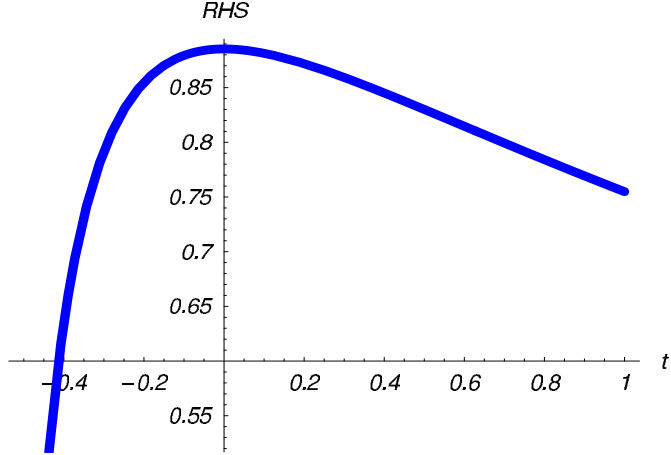}
\caption{Dependence of the RHS of (\ref{4.3b}) on $t$ provided we substitute for $r = -t/(2t+1)$. The minimum is attained for $t= -1/2$ and $t= \infty$
(and hence $r = \infty$ and $r =-1/2$, respectively) while maximum is at $t = 0_+$ (and hence for $r = 0_-$).}%
\label{fig2ab}
\vspace{0.2cm} \hrule
\end{center}
\end{figure}

When the negative binary logarithm is applied to both sides of
(\ref{4.2}), then
\begin{eqnarray}
{\mathcal{I}}_{1+t}({\mathcal{F}}^{(2)}) +
{\mathcal{I}}_{1+r}({\mathcal{F}}^{(1)}) \ \geq\ -D + \frac{1}{r}
\log_2 (1+r)^{D/2} + \frac{1}{t} \log_2 (1+t)^{D/2}\, . \label{4.3}
\end{eqnarray}
Because $1/t + 1/r = -2$, we can recast the previous inequality in the equivalent form
\begin{eqnarray}
{\mathcal{I}}_{1+t}({\mathcal{F}}^{(2)}) +
{\mathcal{I}}_{1+r}({\mathcal{F}}^{(1)}) \ \geq\ \frac{1}{r}
\log_2 [2(1+r)]^{D/2} + \frac{1}{t} \log_2 [2(1+t)]^{D/2}\, . \label{4.3b}
\end{eqnarray}
This can be further simplified by looking at the minimal value of the RHS of (\ref{4.3}) (or (\ref{4.3b})) under the constraint $1/t + 1/r = -2$. The minimal value is attained for $t = -1/2$ and equivalently $t = \infty$, see Fig.~\ref{fig2ab}, and it is $0$ (note $t$ cannot be smaller than $-1/2$).
%
This in particular implies that
%
\begin{eqnarray}
{\mathcal{I}}_{1+t}({\mathcal{F}}^{(2)}) +
{\mathcal{I}}_{1+r}({\mathcal{F}}^{(1)}) \ \geq\ 0\, . \label{4.4}
\end{eqnarray}
Inequality (\ref{4.3}) is naturally stronger
than (\ref{4.4}), but the latter is usually much easier to implement in practical
calculations. In addition the RHS of (\ref{4.3}) is {\em universal} in the sense that it is
$t$ and $r$ independent. The reader should also notice that the zero value
of the right-hand side of (\ref{4.4}) does not yield a trivial inequality since
${\mathcal{I}}_{\alpha}$ are not generally
positive for continuous PDFs (q.v. Section~4.2). In fact, from the coarse probability version of (\ref{4.4})
(cf. Eq.~(\ref{ren6})) follows
\begin{eqnarray}
{\mathcal{I}}_{1+t}({\mathcal{P}}_n^{(2)}) +
{\mathcal{I}}_{1+r}({\mathcal{P}}_n^{(1)}) \ \geq\ - 2D\log_2 l\, , \label{4.4b}
\end{eqnarray}
which is clearly a non-trivial ITUR.

Also, notice that in the limit $t\rightarrow 0_+$ and $r\rightarrow 0_-$ the inequalities (\ref{4.3})-(\ref{4.3b})
take the form~\footnote{Note that these limits do not contradict the constraint $1/t + 1/r = -2$. To see this we go back to the defining equation $1/[2(1+t)] + 1/[2(1+r)] = 1$ which is clearly satisfied for the simultaneous limits $t\rightarrow 0_+$ and $r\rightarrow 0_-$.}
\begin{eqnarray}
{\mathcal{H}}({\mathcal{F}}^{(2)}) +
{\mathcal{H}}({\mathcal{F}}^{(1)}) \ \geq \  \log_2\left(\frac{e}{2} \right)^{D}\, ,
\label{4.5b}
\end{eqnarray}
which coincides with the classical Hirschman conjecture for the differential Shannon entropy based ITUR~\cite{bourret:58}. In this connection it should be noted that among all admissible pairs $\{r,t\}$, the pair $\{0_-,0_+ \}$ gives the highest value of the RHS in (\ref{4.3b}). This can be clearly seen from Fig.~\ref{fig2ab}.

Let us finally observe that when (\ref{4.3})-(\ref{4.3b}) is rewritten in the language of R\'{e}nyi entropy powers it can be equivalently
cast in the form
\begin{eqnarray}
N_{1+t}({\mathcal{F}}^{(2)})N_{1+r}({\mathcal{F}}^{(1)}) \ \equiv \   N_{1+t}({\mathcal{X}})N_{1+r}({\mathcal{Y}})
\ \geq \ \frac{1}{16\pi^2} \, ,
\label{V.59.a}
\end{eqnarray}
or equivalently as
\begin{eqnarray}
N_{p/2}({\mathcal{X}})N_{q/2}({\mathcal{Y}}) \ \geq \ \frac{1}{16\pi^2} \, ,
\label{V.60.a}
\end{eqnarray}
with $p$ and $q$ being H\"{o}lder conjugates and the RE measured in {\em bits}.
Note that in the case when both ${\mathcal{X}}$ and ${\mathcal{Y}}$ represent random Gaussian vectors then (\ref{V.60.a}) reduces to
\begin{eqnarray}
|K_{{\mathcal{X}}}|^{1/D} |K_{{\mathcal{Y}}}|^{1/D} \ = \ \frac{1}{16\pi^2} \, .
\label{V.60.aaa}
\end{eqnarray}
Here, $|K_{{\mathcal{X}}}|$ and $|K_{{\mathcal{Y}}}|$ are determinants of the respective covariance matrices. The equality follows from  the Lieb condition on the saturation of the Beckner--Babenko inequality. It is also interesting to notice that when we define the variance per component, i.e.,
\begin{eqnarray}
&&\sigma_{{\mathcal{X}}}^2 \ = \ \mbox{Var}({\mathcal{X}})/D \ = \ \mbox{Tr}[(K_{{\mathcal{X}}}){_{{ij}}}]/D \, , \\[2mm]
&&\sigma_{{\mathcal{Y}}}^2 \ = \ \mbox{Var}({\mathcal{Y}})/D \ = \ \mbox{Tr}[(K_{{\mathcal{Y}}}){_{{ij}}}]/D \, ,
\end{eqnarray}
then, from (\ref{V.60.aaa}), these satisfy
\begin{eqnarray}
\sigma_{{\mathcal{X}}}^2 \sigma_{{\mathcal{Y}}}^2 \ \geq \ \frac{1}{16\pi^2} \, .
\end{eqnarray}
The proof  is based on the identity
\begin{eqnarray}
\log(\det \mathbb{A}) \ = \ {\mbox{Tr}}( \log  \mathbb{A})\, ,
\end{eqnarray}
which is certainly valid for any diagonalizable matrix $\mathbb{A}$, and more generally for {\em all matrices} since diagonalizable matrices are dense. With this we have
\begin{eqnarray}
\log |K_{{\mathcal{X}}}|^{1/D} \ &=& \ \mbox{Tr}\left[\frac{1}{D} (K_{{\mathcal{X}}}){_{{ij}}} \right]
\ = \ \sum_{i=1}^{D} \left[\frac{1}{D} \log(K_{{\mathcal{X}}}){_{{ii}}} \right] \nonumber \\[3mm]
&\leq& \ \log\left[
\sum_{i=1}^{D} \frac{1}{D} (K_{{\mathcal{X}}}){_{ii}}  \right] \ = \  \log\left\{ \mbox{Tr}[(K_{{\mathcal{X}}}){_{{ij}}}]/D \right\}\ = \ \log \sigma_{{\mathcal{X}}}^2 \, .
\label{V.62.aaa}
\end{eqnarray}
The inequality follows from Jensen's inequality for the logarithm.  An analogous result holds also for the random vector ${\mathcal{Y}}$. The equality in (\ref{V.62.aaa}) holds only when  ${\mathcal{X}}$  is {\em white},  i.e., if its covariance matrix is proportional to
the identity matrix. If the components of the random Gaussian vector are independent, it makes sense (in view of the additivity of the RE) to speak about the RE (and ensuing entropy power) of a given random component. In that case (\ref{V.60.aaa}) boils down to
\begin{eqnarray}
\sigma_{{\mathcal{X}}_i}^2 \sigma_{{\mathcal{Y}}_i}^2 \ = \ \frac{1}{16\pi^2} \, ,
\label{V.67.aaa}
\end{eqnarray}
where the subscript $i$ denotes the $i$-th component of the random vector.

\begin{figure}
[ptb]
\begin{center}
\includegraphics[
height=3.0in, width=4.8in ] {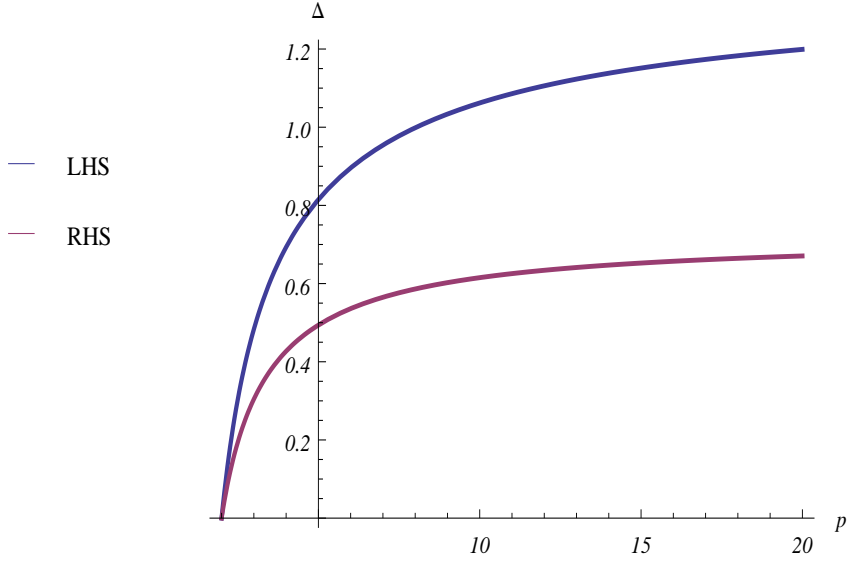} \caption{Graphical representation of the LHS and the RHS of the inequality of (\ref{V.49.ab}). $\Delta$ denotes the difference of entropies on the LHS respective the RHS of (\ref{V.49.ab}). The $q$ variable is phrased in terms of $p$ as $q=p/(p-1)$.  Because the PDF's involved have no fundamental scale, the ensuing entropies (and hence $\Delta$) are $c$ independent. Note in particular that the inequality (\ref{V.49.ab}) is  strongest for $p \rightarrow \infty$ and $q=1$. } \label{fig3b}
\vspace{0.2cm} \hrule
\end{center}
\end{figure}

Inequalities (\ref{V.59.a})-(\ref{V.60.a}) make the connection of the continuous ITUR with the VUR.
This is because when the distributions in question have {\em finite} covariance matrices then the following theorem holds:
\begin{theorem}
Let ${\mathcal{X}}$  be a random vector in $\mathbb{R}^D$ with the finite covariance matrix $(K_{{\mathcal{X}}}){_{{ij}}}$. Then
\begin{eqnarray}
N({\mathcal{X}}) \ \leq \  |K_{{\mathcal{X}}}|^{1/D} \ \leq \ \sigma^2_{{\mathcal{X}}}\, ,
\end{eqnarray}
with equality in the first inequality if and only if  ${\mathcal{X}}$ is a Gaussian vector, and
in the second if and only if ${\mathcal{X}}$ is white.

\end{theorem}
The proof of this theorem is based on the non-negativity of the relative Shannon entropy
(or Kullback--Leibler divergence) and can be found, e.g., in Refs.~\cite{Dembo:91,Rioul:11}. An important upshot of the previous
theorem is that also for {\em non-Gaussian} distributions one has
\begin{eqnarray}
\sigma_{{\mathcal{X}}}^2 \sigma_{{\mathcal{Y}}}^2 \ \geq \ |K_{{\mathcal{X}}}|^{1/D} |K_{{\mathcal{Y}}}|^{1/D} \ \geq \ N({\mathcal{X}})N({\mathcal{Y}}) \ \geq \ \frac{1}{16\pi^2} \, ,
\label{V.48.ab}
\end{eqnarray}
which saturates only for Gaussian (respective white) random vectors ${\mathcal{X}}$  and ${\mathcal{Y}}$. This is just one
example where a well-known inequality can be improved by replacing variance
by a quantity related to entropy. In certain cases the last inequality in (\ref{V.48.ab}) can be
improved by using R\'{e}nyi's entropy power rather than Shannon's entropy power. In fact,
the inequality
\begin{eqnarray}
 N({\mathcal{X}})N({\mathcal{Y}}) \ \geq \ N_{p/2}({\mathcal{X}})N_{q/2}({\mathcal{Y}})  \ \geq \ \frac{1}{16\pi^2} \, ,
\end{eqnarray}
(with $p$ and $q$ being H\"{o}lder conjugates) is fulfilled whenever
\begin{eqnarray}
\mathcal{H}({\mathcal{X}}) - \mathcal{I}_{p/2}({\mathcal{X}}) \ \geq \ \mathcal{I}_{q/2}({\mathcal{Y}}) - \mathcal{H}({\mathcal{Y}})\, ,
\label{V.49.ab}
\end{eqnarray}
($q\in [1,2]$ and $p\in [2, \infty)$). Note that both sides in
(\ref{V.49.ab}) are positive (cf. Eq.~(\ref{ren6})).
Inequality~(\ref{V.49.ab})  can be satisfied by a number of PDFs.
This is often the case when the PDF ${\mathcal{F}}^{(1)}$ associated
with ${\mathcal{Y}}$ is substantially {\em leptokurtic} (peaked)
while ${\mathcal{F}}^{(2)}$ (which is related to ${\mathcal{X}}$) is
{\em platykurtic} heavy-tailed PDF. A simple example is the Cauchy--Lorentz distribution~\cite{feller}, for which
we have
\begin{eqnarray}
&&f(x) \ = \ \sqrt{\frac{c}{\pi}} \ \! \sqrt{\frac{1}{c^2 + x^2}}\, , \\[2mm]
&&\hat{f}(y) \ = \  \sqrt{\frac{2c}{\pi^2  }}\ \! K_0(c |y|)\, ,\\[2mm]
&&\mathcal{F}^{(2)}(x) \ = \ \frac{c}{\pi} \ \! {\frac{1}{c^2 + x^2}}\, , \label{73ab}\\[2mm]
&&\mathcal{F}^{(1)}(y) \ = \ {\frac{2c}{\pi^2 }}\ \! K_0^2(c |y|)\, .\label{73abc}
\end{eqnarray}
Here $\mathcal{F}^{(2)}(x) $ is the Cauchy--Lorentz PDF. In Fig~\ref{fig3b} we graphically
represent the LHS and the
RHS of the inequality of (\ref{V.49.ab}) for  PDFs (\ref{73ab})-(\ref{73abc}).
There we can see that the inequality becomes strongest if we choose $p \rightarrow \infty$ and $q=1$.
On the basis of our numerical
simulations it seems that the behavior depicted in Fig~\ref{fig3b} is quite common for
heavy-tailed L\'{e}vy stable distributions.

Of course, in cases when the covariance matrices are infinite, Theorem~V.2 does not hold
and the continuous ITUR remains the only sensible quantifier of the inherent QM uncertainty.
We shall delve more  into the quantum mechanics implications of the above (continuous PDF) ITUR
inequalities in Section~\ref{SEc5.b}.


\section{Applications in Quantum Mechanics\label{SEc5}}

The connection of the information theoretic inequalities
(\ref{3.3.1}) and (\ref{4.3}) with quantum mechanics is established
when we consider two quantum-mechanical  observables, say $\hat{A}$
and $\hat{B}$, written through their spectral decompositions
\begin{eqnarray}
\hat{A} = \sum\hspace{-5mm}\int a |a\rangle\langle a| \ da \, , \;\;\;\;\;\;\;\;
\hat{B} = \sum\hspace{-5mm}\int b |b\rangle\langle b| \ db \, .
\end{eqnarray}
Here the integral-summation symbol schematically represents summation over a discrete part of the spectra
and integration over a continuous part of the spectra. States $|a \rangle $ and $| b \rangle$ are proper
(for discrete spectrum) and improper (for the continuous spectrum) eigenvectors of $\hat{A}$
and $\hat{B}$, respectively.

According to the quantum measurement postulate, the  probability of obtaining a
result $a$ in a measurement of observable $\hat{A}$ on a system prepared in the state $|\phi\rangle$
is given by the (transition) probability density
\begin{eqnarray}
{\mathcal{F}}(a) = | \langle a | \phi \rangle |^2\, .
\end{eqnarray}
When $a_i$ belongs to a discrete spectrum, then the (transition)
probability for the result $a_i$  is
\begin{eqnarray}
p(a_i) = | \langle a_i | \phi \rangle |^2\, .
\end{eqnarray}
Similarly for the observable $\hat{B}$.

\subsection{Discrete probabilities\label{SEc5.a}}

\begin{figure}
[ptb]
\begin{center}
\includegraphics[
height=2.3601in, width=5.6662in ]{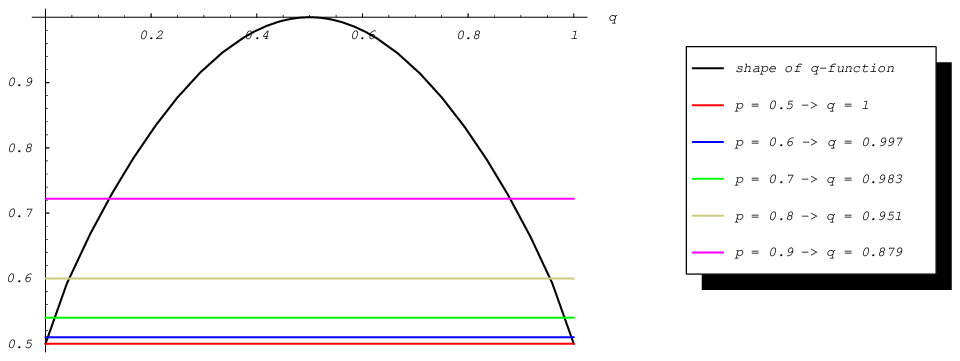} \caption{Graphical representation of the inequality (\ref{6.46ab}). For fixed value of $p$ the inequality is fulfilled for all $q$'s that lie inside the $q$-function, i.e., function $y = \mbox{$\frac{1}{2}$}\ \! q^{-q}(1-q)^{q-1}$.} \label{fig3aa}
\vspace{0.2cm} \hrule
\end{center}
\end{figure}

\begin{figure}
[ptb]
\begin{center}
\includegraphics[
height=2.3601in, width=5.6662in ] {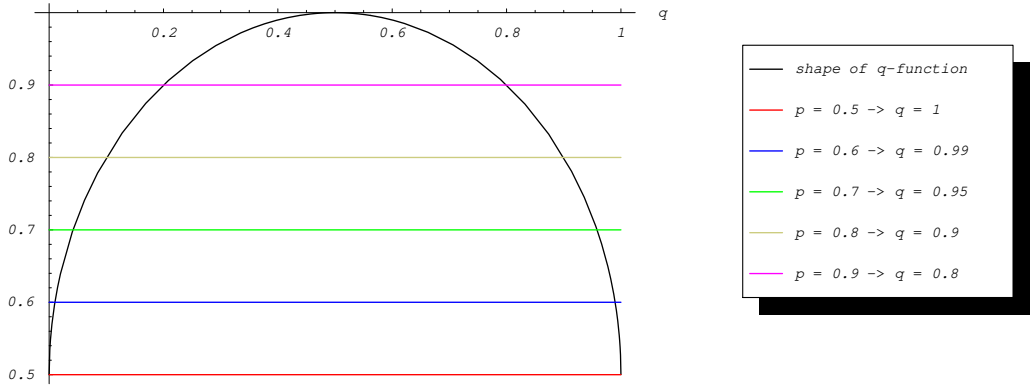} \caption{Graphical representation of the inequality (\ref{6.48a}). For fixed value of $p$ the inequality is fulfilled for all $q$'s that lie inside the $q$-function, i.e., function $y = \sqrt{q}\sqrt{1-q} + 1/2$.} \label{fig3a}
\vspace{0.2cm} \hrule
\end{center}
\end{figure}

For the discrete-spectrum the conditions
assumed in the Riesz--Thorin inequality (cf. Theorem~IV.2) are clearly
fulfilled by setting $x_i = \langle x_i| \phi \rangle$, $(\mathcal{L} {\bf x})_j = \langle b_j| \phi \rangle$ and $a_{ij} = \langle b_j|a_k \rangle$. We will now illustrate the utility of R\'{e}nyi's ITUR with a toy-model example.
To this end we consider a two-dimensional state $|\phi\rangle$ of a
spin-$\frac{1}{2}$ particle, and let $\hat{A}$ and $\hat{B}$ be {\em
spin components} in {\em orthogonal} directions, i.e.
\begin{eqnarray}
|A\rangle \equiv \left(
                      \begin{array}{c}
                        |S_x;+\rangle \\
                        |S_x;-\rangle  \\
                      \end{array}
                    \right), \;\;\;\;\;\;
|B\rangle \equiv \left(
                      \begin{array}{c}
                        |S_z;+\rangle \\
                        |S_z;-\rangle  \\
                      \end{array}
                    \right).
\end{eqnarray}
Because
\begin{eqnarray}
\left(
                      \begin{array}{c}
                        |S_x;+\rangle \\
                        |S_x;-\rangle  \\
                      \end{array}
                    \right)  \ = \
\frac{1}{\sqrt{2}}\left(
  \begin{array}{cc}
    {1} & {1} \\
    -{1} & {1}\\
  \end{array}
\right) \left(
                      \begin{array}{c}
                        |S_z;+\rangle \\
                        |S_z;-\rangle  \\
                      \end{array}
                    \right)\, ,
\end{eqnarray}
we can immediately identify  $c$ with ${1}/{\sqrt{2}}$  (cf. Eq.~(\ref{kendt})).
Let us now define probability ${\mathcal{P}} = (p,(1-p)) \equiv (|\langle
S_x;+|\phi\rangle|^2,|\langle S_x;-|\phi\rangle|^2)$. Without loss of generality we may assume that $p = \max_i{{\mathcal{P}}}$. The question
we are interested in is how the knowledge of ${\mathcal{P}}$ restricts the distribution
${\mathcal{Q}} = (q,(1-q)) \equiv (|\langle S_z;+|\phi\rangle|^2,|\langle S_z;-|\phi\rangle|^2)$? Both distribution cannot be independent as Shannon's ITUR
\begin{eqnarray}
{\mathcal{H}}({\mathcal{P}}) \ + \
{\mathcal{H}}({\mathcal{Q}}) \ \geq \ -
2\log_2 c \ = \ 1\, ,
\label{6.45}
\end{eqnarray}
clearly indicates. In fact, inequality (\ref{6.45}) can be equivalently phrased in the form
\begin{eqnarray}
p^p(1-p)^{1-p} \ \leq
\ \mbox{$\frac{1}{2}$}\ \! q^{-q}(1-q)^{q-1}\, .
\label{6.46ab}
\end{eqnarray}
The graphical solution of this equation can be seen on Fig.~\ref{fig3aa}.
In case of R\'{e}nyi's ITUR we can take advantage of the fact that the RE is a monotonically
decreasing function of its
index and hence the most stringent relation between $p$ and $q$ is provided via R\'{e}nyi's ITUR
\begin{eqnarray}
&&\mathcal{I}_{\infty}({\mathcal{P}}) \ + \
\mathcal{I}_{1/2}({\mathcal{Q}})\
\geq\ -2\log_{2}c  \ = \ 1 
\, .
\end{eqnarray}
The latter is equivalent to
\begin{eqnarray}
\sqrt{q}\sqrt{1-q} + 1/2 \ \geq \ p
\, .
\label{6.48a}
\end{eqnarray}
%
This inequality can be again treated graphically, see Fig.~\ref{fig3a}.
The comparison with the ordinary Schr\"{o}dinger--Robertson's VUR,
can easily be made. In fact, we have
%
%
\begin{eqnarray}
&&\mbox{\hspace{-4mm}}{\langle (\triangle S_x)^2\rangle_{\phi}
\langle (\triangle S_z)^2 \rangle_{\phi} \ \geq \
\frac{\hbar^2}{4}|\langle S_y\rangle_{\phi}|^2}\;\;\;\Leftrightarrow
\;\;\; {p(1-p) \ \geq \ \mbox{$\frac{1}{4}$} \sin^2(\varphi_+ -
\varphi_-) }\, ,
\label{VI.A.84.a}
\end{eqnarray}
%
where the phase $\varphi_{\pm}$ is defined as
\begin{eqnarray}
e^{i\varphi_{\pm}} \ \equiv \  \frac{\langle \phi| S_z;
\pm\rangle}{|\langle \phi| S_z; \pm\rangle|} \, .
\end{eqnarray}
In deriving this we have used the relation
\begin{eqnarray}
\left(
                      \begin{array}{c}
                        |S_y;+\rangle \\
                        |S_y;-\rangle  \\
                      \end{array}
                    \right)  \ = \
\frac{1}{\sqrt{2}}\left(
  \begin{array}{cc}
    {1} & {i} \\
    {1} & -{i}\\
  \end{array}
\right) \left(
                      \begin{array}{c}
                        |S_z;+\rangle \\
                        |S_z;-\rangle  \\
                      \end{array}
                    \right) .
\end{eqnarray}
Note, that by symmetry the VUR inequality can also equally be
written as
\begin{eqnarray}
{q(1-q) \ \geq \ \mbox{$\frac{1}{4}$} \sin^2(\tilde{\varphi}_+ -
\tilde{\varphi}_-) }\, ,
\label{VI.A.87.a}
\end{eqnarray}
with
\begin{eqnarray}
e^{i\tilde{\varphi}_{\pm}} \ \equiv \  \frac{\langle \phi| S_x;
\pm\rangle}{|\langle \phi| S_x; \pm\rangle|} \, .
\end{eqnarray}

From (\ref{VI.A.84.a}) and (\ref{VI.A.87.a}) we see that the VUR does not
pose any strong restriction between ${\mathcal{P}}$ and ${\mathcal{Q}}$.  Since the phase
factors $\varphi_{\pm}$ (or $\tilde{\varphi}_{\pm}$) do not enter the definition
of ${\mathcal{Q}}$ (or ${\mathcal{P}}$), then for a fixed (but otherwise arbitrary) $q$ the VUR
(\ref{VI.A.84.a}) can be in principle fulfilled by  any $p\in[0.5,1]$. Of course, if the relative phase is known
the restriction between ${\mathcal{P}}$ and ${\mathcal{Q}}$ is less trivial.  On the other hand, the ITURs discussed above
are far more specific in their constrains on values of ${\mathcal{P}}$ and ${\mathcal{Q}}$, see Tab.~\ref{tab.1}.
\begin{table}[htb]
\centering
\begin{tabular}{|c|c|c|c|}
 \hline
$~p~$&~VUR $q\in$~&~S-ITUR $q\in$~&~R-ITUR $q\in~$\\ \hline \hline
  $~0.5~$&$~[0.067,0.933]~$&$~[0,1]~$&$~[0,1]~$\\
  $~0.6~$&$~[0.067,0.933]~$&$~[0.003,0.997]~$&$~[0.010,0.990]~$\\
  $~0.7~$&$~[0.067,0.933]~$&$~[0.017,0.983]~$&$~[0.042,0.958]~$\\
  $~0.8~$&$~{[0.067,0.933]}~$&$~{[0.049,0.951]}~$&$~{[0.1,0.9]}~$\\
  $~0.9~$&$~{[0.067,0.933]}~$&$~{[0.121,0.879]}~$&$~{[0.2,0.8]}~$ \\
  \hline
\end{tabular}
\caption{Comparison of three uncertainty relations: variance-based uncertainty relation (VUR) with $\tilde{\varphi}_+ -
\tilde{\varphi}_- = \pi/6$,
Shannon's information uncertainty relation (S-ITUR) and R\'{e}nyi's information uncertainty relation (R-ITUR)
for different values of $p$. In the respective columns one can see the peakedness of the distribution
${\mathcal{Q}} = (q,(1-q))$.}\label{tab.1}
\vspace{0.2cm} \hrule
\end{table}
From the table we see that for given $\mathcal{P}$, R\'{e}nyi's ITUR improves on Shannon's ITUR. This is because the R\'{e}nyi ITUR considered
is more restrictive than Shannon's case. For instance, the marginal case $\mathcal{P} = (0.8,0.2)$ and
$\mathcal{Q} = (0.951, 0.049)$ that is allowed by
Shannon's ITUR explicitly violates R\'{e}nyi's ITUR and hence it cannot be realized (ITURs represent necessary conditions).
%
%
%
Both Shannon's ITUR and R\'{e}nyi's ITUR  improve on VUR --- unless some extra information about the relative wave-functions
phase is provided. In Tab.~\ref{tab.1} we find that when the relative phase is known, e.g., $\tilde{\varphi}_+ -
\tilde{\varphi}_- = \pi/6$,  R\'{e}nyi's ITUR still improves on VUR for values $p = 0.9$ and $p=0.8$ while Shannon's ITUR
improves over VUR only for $p=0.9$.
\\[5mm]


\subsection{Continuous probabilities\label{SEc5.b}}

In view of the (continuous) ITUR from Section~\ref{SEc4}  the most prominent example of the Fourier transform is that
between configuration and momentum space wave functions (analogously one can treat also other Fourier transform duals,
such as the angular momentum and angle). In particular between $\psi(\bf{x})$ and $\hat{\psi}(\bf{p})$ hold two reciprocal relations
\begin{eqnarray}
&&\psi({\bf{x}})  \ = \ \int_{\mathbb{R}^D} e^{i {\bf p}\cdot {\bf x}/\hbar} \  \! \hat{\psi}({\bf{p}})\ \!  \frac{d{\bf p}}{(2\pi
\hbar)^{D/2}}\, ,\nonumber \\[2mm]
&&\hat{\psi}({\bf{p}})  \ = \ \int_{\mathbb{R}^D} e^{-i {\bf p}\cdot {\bf x}/\hbar} \  \! {\psi}({\bf{x}})\ \!  \frac{d{\bf x}}{(2\pi
\hbar)^{D/2}}\, . \label{V.1.a}
\end{eqnarray}
The Plancherel (or Riesz--Fischer) equality~\cite{Hardy1959,Yosida1968} then implies that $|\!|{\psi}|\!|_2 = |\!|
\hat{\psi}|\!|_{2} = 1$. Let us define new functions in (\ref{V.1.a}), namely
\begin{eqnarray}
&&f^{(2)}({\bf{x}}) \ = \ (2\pi \hbar)^{D/4}\psi(\sqrt{2\pi\hbar}\ \! {\bf{x}})\, , \nonumber \\[2mm]
&&f^{(1)}({\bf{p}}) \ = \ (2\pi \hbar)^{D/4}\hat{\psi}(\sqrt{2\pi\hbar}\ \! {\bf{p}}) \, .
\label{6.2.a}
\end{eqnarray}
The factor $(2\pi \hbar)^{D/4}$ ensures that also the new functions are normalized (in sense of $|\!|\ldots|\!|_2$) to unity.
With these we will have the same structure of the Fourier transform  as in the Beckner--Babenko theorem in Section~\ref{SEc4}.
Consequently we can write the associated ITURs (\ref{4.3})-(\ref{4.3b}) in the form
\begin{eqnarray}
{\mathcal{I}}_{1+t}(|{\psi} |^2) +
{\mathcal{I}}_{1+r}(|\hat{\psi} |^2) \ &\ge& \ D\log_2(\pi\hbar) + \frac{1}{r}
\log_2 (1+r)^{D/2} + \frac{1}{t} \log_2 (1+t)^{D/2}\nonumber \\[2mm]
&=& \ \frac{1}{r}
\log_2 \left(\frac{1+r}{\pi\hbar}\right)^{\!D/2} + \frac{1}{t} \log_2 \left(\frac{1+t}{\pi \hbar}\right)^{\!D/2}\, , \label{6.3}
\end{eqnarray}
or in the weaker form with the universal RHS
\begin{eqnarray}
{\mathcal{I}}_{1+t}(|{\psi} |^2) +
{\mathcal{I}}_{1+r}(|\hat{\psi} |^2) \ \geq\ \log_2(2\pi\hbar)^D \, . \label{6.3.b}
\end{eqnarray}
In particular for Shannon's entropy  the Hirschman inequality (\ref{4.5b}) acquires the form
\begin{eqnarray}
{\mathcal{H}}(|{\psi} |^2) +
{\mathcal{H}}(|\hat{\psi} |^2) \ \geq \  \log_2(e\pi\hbar)^D\, .
\label{71.abc}
\end{eqnarray}
In both (\ref{6.3}) and (\ref{6.3.b}) use was made of the mathematical identities
\begin{eqnarray}
&&\mathcal{I}_{\alpha}(|f^{(1)}|^2) \ = \ \mathcal{I}_{\alpha}(|\hat{\psi} |^2) - \frac{D}{2} \log_2 (2\pi \hbar)\, ,\nonumber \\[2mm]
&&\mathcal{I}_{\alpha}(|f^{(2)}|^2) \ = \ \mathcal{I}_{\alpha}(|{\psi} |^2) - \frac{D}{2} \log_2 (2\pi \hbar)\, .
\end{eqnarray}
These two identities just state that the scaled PDFs $|f^{(1)}|^2$ and $|f^{(2)}|^2$ obtained from (\ref{6.2.a}) are less peaked (and hence less informative) than the original PDFs $|\hat{\psi}|^2$ and $|{\psi}|^2$, respectively. Consequently, we increase our ignorance when passing from $\hat{\psi}$ to $f^{(1)}$, and from ${\psi}$ to $f^{(2)}$.

The inequality (\ref{6.3.b}) (and similarly (\ref{6.3})) should be understood in the sense that by no quantum mechanical measurements it is possible to reduce the joint entropy in two canonically conjugate distributions $\mathcal{F}^{(1)}({\bf{p}}) = |\hat{\psi}({\bf{p}})|$  and $\mathcal{F}^{(2)}({\bf{x}}) = |{\psi}({\bf{x}})|$ below the level of $\log_2(2\pi\hbar)^D$ bits.

Let us observe that in terms of the R\'{e}nyi entropy power one can cast (\ref{6.3}) into an equivalent form (cf. Eq.~(\ref{V.60.a}))
\begin{eqnarray}
N_{1+t}(|\psi|^2)N_{1+r}(|\hat{\psi}|^2) \ \geq \ \frac{\hbar^2}{4} \, .
\end{eqnarray}
%

\subsubsection{Heavy tailed distributions\label{SEc5.bab}}

If we wish to improve over the Shannon--Hirschman ITUR (\ref{4.5b}) we should find such a pair $\{r,t\}$ which provides a stronger restriction on the involved distributions than
Shannon's case. In Section~\ref{SEc4} we have already seen that this can indeed happen, e.g., for heavy tailed distributions. This fact will be now illustrated with {\em Paretian} or {\em L\'{e}vy (stable)} distributions. Such distributions represent, in general,  a four parametric class of distributions that replace the r\^{o}le of the {\em normal distribution} in the central limit theorem in cases where the underlying single event distributions do not have one of the first two momenta. For computational simplicity (results can be obtained in a closed form) we will consider one of the  L\'{e}vy stable distributions, namely the Cauchy--Lorentz distribution~\cite{feller} which can be obtained from the wave function
\begin{eqnarray}
\hat{\psi}(x) \ = \ \sqrt{\frac{c}{\pi}} \ \! \sqrt{\frac{1}{c^2 + (x-m)^2}}\, .
\end{eqnarray}
The corresponding Fourier transform and respective PDFs are
\begin{eqnarray}
&&\hat{\psi}(p) \ = \ e^{-imp/\hbar}\ \! \sqrt{\frac{2c}{\pi^2 \hbar }}\ \! K_0(c |p|/\hbar)\, ,\\[2mm]
&&\mathcal{F}^{(2)}(x) \ = \ \frac{c}{\pi} \ \! {\frac{1}{c^2 + (x-m)^2}}\, , \\[2mm]
&&\mathcal{F}^{(1)}(p) \ = \ {\frac{2c}{\pi^2 \hbar }}\ \! K_0^2(c |p|/\hbar)\, ,
\end{eqnarray}
and the ensuing Shannon and R\'{e}nyi entropies are
\begin{eqnarray}
&&\mathcal{H}(\mathcal{F}^{(1)}) \ = \ \log_2(\pi^2 \hbar/2c) -\frac{8}{\pi^2}\ \! 2.8945\, , \;\;\;  \mathcal{H}(\mathcal{F}^{(2)}) \ = \ \log_2(4c\pi)\, , \nonumber \\[2mm]
&&\mathcal{I}_{1/2}(\mathcal{F}^{(1)}) \ = \ \log_2(2 \hbar/c )\, , \;\;\;  \mathcal{I}_{\infty}(\mathcal{F}^{(2)}) \ = \ \log_2(c\pi)\, .
\end{eqnarray}
With these results we can immediately write the associated ITURs, namely
\begin{eqnarray}
&&{\mathcal{H}}(\mathcal{F}^{(1)}) +
{\mathcal{H}}(\mathcal{F}^{(2)}) \ = \ \log_2(2\pi^3 \hbar) -  \frac{8}{\pi^2} \ \! 2.8945 \ > \  \log_2(e\pi \hbar)\, ,\label{78.aa}\\[3mm]
&&{\mathcal{I}}_{1/2}(\mathcal{F}^{(1)}) +
{\mathcal{I}}_{\infty}(\mathcal{F}^{(2)}) \ = \  \log_2(2\pi \hbar)\, .\label{79.aa}
\end{eqnarray}
So what can be concluded from these relations? First we notice that the ITUR (\ref{79.aa}) saturates the inequality (\ref{6.3}) while the Shannon ITUR (\ref{78.aa}) does not saturate the corresponding Hirschman inequality (\ref{71.abc}). In fact, if we rewrite (\ref{78.aa})-(\ref{79.aa}) in the language of R\'{e}nyi entropy powers, we obtain
\begin{eqnarray}
&&N(\mathcal{F}^{(1)})N(\mathcal{F}^{(2)}) \ > \ \frac{\hbar^2}{4} \, ,\label{78.aaa}\\[3mm]
&&N_{1/2}(\mathcal{F}^{(1)})N_{\infty}(\mathcal{F}^{(2)}) \ = \ \frac{\hbar^2}{4}\, .\label{79.aaa}
\end{eqnarray}
Since the R\'{e}nyi ITUR  puts a definite constraint between $\mathcal{F}^{(2)}$ and $\mathcal{F}^{(1)}$  it clearly improves over the Shannon ITUR (which  is less specific). In addition, while (\ref{78.aaa}) indicates that one could still find another $\mathcal{F}^{(2)}$ for a given fixed $\mathcal{F}^{(1)}$ that would lower the LHS of the Shannon entropy power inequality, the relation (\ref{79.aaa}) forbids such a situation to happen without increasing uncertainty in the R\'{e}nyi ITUR. By increasing the uncertainty, however, the definite constraint between  $\mathcal{F}^{(2)}$ and $\mathcal{F}^{(1)}$ will get lost.

It should be stressed, that in general the R\'{e}nyi ITUR is not symmetric.
However, in the case at hand the situation is quite interesting. One can easily check that ${\mathcal{I}}_{1/2}(\mathcal{F}^{(2)}) = \infty$ and ${\mathcal{I}}_{\infty}(\mathcal{F}^{(1)}) = -\infty$, and so the
R\'{e}nyi ITUR is indeterminate. This result deserves two comments. First, the extremal values of ${\mathcal{I}}_{1/2}(\mathcal{F}^{(2)})$ and
${\mathcal{I}}_{\infty}(\mathcal{F}^{(1)})$ can be easily understood. From the very formulation of the RE one can see that for $\alpha >1$ the non-linearly nature of the RE tends to  emphasize the more probable parts of the PDF (typically the middle parts) while for $\alpha <1$) the less probable parts of the PDF (typically the tails) are accentuated. In other words,  ${\mathcal{I}}_{1/2}$ mainly carries information on the rare events while ${\mathcal{I}}_{\infty}$ on the common events. In particular, if one starts from a strongly leptocurtic distribution (such as $\mathcal{F}^{(1)}$) then ${\mathcal{I}}_{\infty}$ effectively works with the PDF that is sharply (almost $\delta$-function) peaked. In this respect ignorance about the peak is minimal, which in turn corresponds to the minimal RE which for continuous distributions is $-\infty$. For heavy tailed distributions (such as $\mathcal{F}^{(2)}$) the RE ${\mathcal{I}}_{1/2}$ works effectively with a very flat (almost equiprobable) PDF which yields maximal ignorance about the tail. For continuous distributions the related information of the order $\alpha =1/2$ is thus $\infty$.

Second,
one can make sense of the indeterminate form of the
R\'{e}nyi ITUR by putting a regulator on the real $x$ axis. In particular we can assume that
$\int_{-\infty}^{\infty} dx \ldots \mapsto \int_{-R}^{R} dx \ldots$. With this we obtain to leading order in $R$
\begin{eqnarray}
&&\mathcal{I}_{1/2}(\mathcal{F}^{(2)}) \ = \ 2 \log_2 \left(\sqrt{\frac{c}{\pi}}\log(4 R^2/c^2) \right)
, \nonumber \\[2mm]
&&\mathcal{I}_{\infty}(\mathcal{F}^{(1)}) \ = \ -\log_2\left(\frac{2 c}{\hbar \pi^2} K_{0}^2(c/R)\right)
.
\end{eqnarray}
In the associated ITUR the unwanted divergent terms cancel and we end up with the final result
\begin{eqnarray}
{\mathcal{I}}_{1/2}(\mathcal{F}^{(2)}) +
{\mathcal{I}}_{\infty}(\mathcal{F}^{(1)}) \ \stackrel{R\rightarrow \infty}{=} \  \log_2(2\pi \hbar)\, ,
\end{eqnarray}
which again, rather surprisingly, saturates the information bound.

It is also interesting to observe that  while the variance in momentum $\langle(\triangle p)^2\rangle_{\psi} = \hbar^2\pi/16c^2$, the variance in position  $\langle(\triangle x)^2\rangle_{\psi} = \infty$ (which is symptomatic of L\'{e}vy stable distributions) and hence the Schr\"{o}dinger--Robertson VUR is completely uninformative. Similar conclusions can be also reached with the L\'{e}vy--Smirnov distribution which is used in fractional QM~\cite{Al-Saqabi:13,Laskin:00} and which can be obtained from the wave function
\begin{eqnarray}
\psi(x) \ = \ \left(\frac{c}{2\pi}\right)^{1/4} \exp\left(-\frac{c}{4}(x-m)^{-1} + \frac{i}{\hbar}\ \!p_{0}x\right)/(x-m)^{3/4}\, .
\end{eqnarray}

Let us finally note that the meaning of the ITUR (\ref{6.3}) (and (\ref{6.3.b})) is rather different from the momentum-position VUR. The difference is due to the fact that the two measures of uncertainty (namely variance and R\'{e}nyi's entropy) are left unaltered by very different types of PDF modifications.
While both the variance of a probability distribution and R\'{e}nyi entropy are translation invariant (i.e., invariant under the shift of the mean value of the distribution by a constant), R\'{e}nyi entropy is, in addition, invariant under the piecewise reshaping of the wave function. Particularly PDF's $\varrho^{(2)}(x) = |\psi^{(1)}( x)|^2$ and $\bar{\varrho}^{(2)}(x)  =  |\bar{\psi}^{(2)}(x)|^2$
with the wave function
\begin{eqnarray}
\bar{\psi}^{(2)}(x) \ = \ \sum_{n\in \mathbb{N}}\chi_{[ndx, (n+1) dx]}\ \!\psi^{(2)}(x_{\sigma(n)})\, ,
\end{eqnarray}
($\chi_{[a,b]}$ is the indicator function of the interval $[a,b]$ in
$\mathbb{R}$ and $\sigma(n)$ is an arbitrary permutation of the set
of all $n \in \mathbb{N}$) yield the same R\'{e}nyi entropy. In
other words, R\'{e}nyi entropy is invariant under cutting up the
original PDF $\varrho^{(2)}(x)$ into infinitesimal pieces under the
original curve and {\em reshuffling} or  {\em separating} them in an
arbitrary manner. Also the R\'{e}nyi entropy for corresponding
Fourier transformed wave functions are unchanged when passing from
$\varrho^{(1)}({\bf p})$ to  $\bar{\varrho}^{(1)}({\bf p})$. This
indicates that the corresponding ITUR will not change under such a
reshuffling.
This fact will be illustrated in the following subsection.

\subsubsection{Schr\"{o}dinger cat states\label{SEc5.bac}}

Another relevant situation when the continuous ITUR improves on the VUR
occurs for coherent state superpositions (CSS), also called Schr\"odinger cat states. These states have the form  %
\begin{eqnarray}
|CSS_{\pm} (\beta) \rangle \ = \  {N}^{\pm}_{\beta} \left(|\beta \rangle \ \pm \ |{\rm -} \beta \rangle\right) \label{CSS1}
\end{eqnarray}
where $|\beta \rangle$ is the ordinary Glauber coherent
state with the amplitude $\beta$ and
\begin{eqnarray}
{N}^{\pm}_{\beta} \ = \ 1 / \sqrt{2(1 \pm {\rm
e}^{-2\beta^2})}\, ,
\end{eqnarray}
is the normalization factor.  Such states have been created in the laboratory~\cite{Brune1996} and are of interest in studies of the quantum to classical transition as well as quantum metrology \cite{Knott2013, Knott2014}. For definiteness we shall consider only the
$|CSS_{+} (\beta) \rangle$  state, though the qualitative statements
will equally hold also for $|CSS_{-} (\beta) \rangle$.   The operator corresponding to different phase quadratures of this state is
\begin{eqnarray}
\hat{X}_{\theta} = (\hat{b} e^{-i\theta} + \hat{b}^\dag e^{i\theta})/2,
\end{eqnarray}
where $\hat{b}^\dag$ and $\hat{b}$ are respectively the creation and annihilation operators for a photon in the coherent state mode. Note that the eigenvalues of these operators are unitless and do not depend on $\hbar$ as was the case with the other examples. We shall be concerned with
the orthogonal quadratures $\hat{X}_{0}$ and  $\hat{X}_{\pi/2}$, which form a pair of conjugate observables with the commutation relation $[\hat{X}_{0}, \hat{X}_{\pi/2}]=i/2$.
If we take $|x_0\rangle$ and $|x_{\pi/2}\rangle$ to be eigenstates of $\hat{X}_{0}$ and $\hat{X}_{\pi/2}$ we can represent (\ref{CSS1}) in these bases as
\begin{eqnarray}
\langle x_0 |CSS_{+} (\beta) \rangle \ &=& \
{N}^{+}_{\beta} \left( \langle x_0 | \beta \rangle \  + \  \langle x_0 | - \beta \rangle \right)\nonumber \\[2mm]
&=& \ \frac{2{N}^{+}_{\beta}}{ \pi^{{1\over 4}}}  \cosh \left(\sqrt{2} \beta x _0\right) \exp \left[ -{1\over 2} x_0^2 - \beta^2 \right] \, ,
\label{Homodyne_03} \\[3mm]
\langle x_{\pi/2} |CSS_{+} (\beta) \rangle  \ &=& \ {N}^{+}_{\beta} \left( \langle x_{\pi/2}  | \beta \rangle \ +
\ \langle x_{\pi/2}  | -\beta \rangle \right) \nonumber \\[2mm]
&=& \ \frac{2{N}^{+}_{\beta}}{\pi^{{1\over 4}}} \cos \left(\sqrt{2}
 \beta x_{\pi/2}  \right) \exp \left[ -{1\over 2} x_{\pi/2} ^2 \right] . \label{Homodyne_04}
\end{eqnarray}
The corresponding probability distributions
\begin{eqnarray}
\mathcal{F}^{(2)}(x_0) &=& \ \langle x_0 |CSS_{+} (\beta) \rangle\langle CSS_{+} (\beta) | x_0 \rangle \\
\mathcal{F}^{(1)}(x_{\pi/2} ) &=& \ \langle x_{\pi/2}  |CSS_{+} (\beta) \rangle\langle CSS_{+} (\beta) | x_{\pi/2}  \rangle\,
\end{eqnarray}
can be experimentally accessed with homodyne detections.

\begin{figure}
[ptb]
\begin{center}
\rotatebox{0}{\includegraphics[
height=5in, width=5in ] {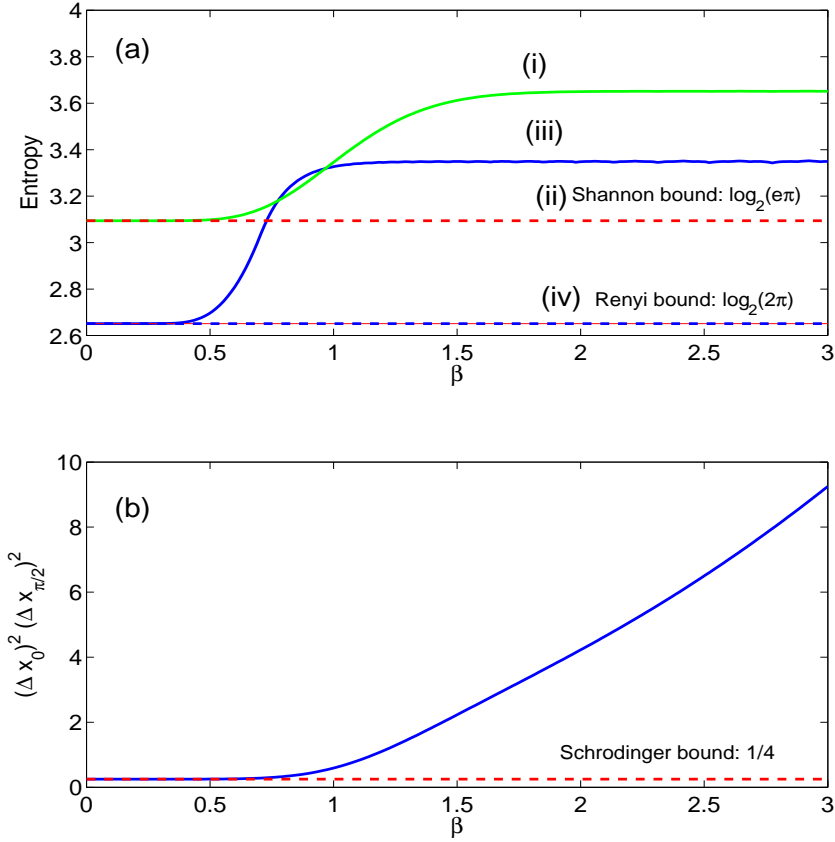}}
\caption{{\bf a)} Plot of different entropies for $|CSS_+(\beta)\rangle$ as a function of $\beta$. (i) Shannon ITUR,  $\mathcal{H}(\mathcal{F}^{(2)}) + \mathcal{H}(\mathcal{F}^{(1)})$, (ii) bound for the Shannon ITUR, $\log_2(e\pi)$ (iii) Renyi ITUR, $\mathcal{I}_{1/2}({\mathcal{F}^{(1)}}) + \mathcal{I}_{\infty}({\mathcal{F}^{(2)}})$, and (iv) the other way round, $\mathcal{I}_{1/2}({\mathcal{F}^{(2)}}) + \mathcal{I}_{\infty}({\mathcal{F}^{(1)}})$. Also shown as a faint solid line overlapping with (iv) is the R\'{e}nyi ITUR bound, $\log_2(2\pi)$.
{\bf b)} Plot of the Robertson--Schr\"{o}dinger VUR for $|CSS_+(\beta)\rangle$ as a function of $\beta$ (solid curve) and its bound (dashed curve).}
 \label{fig6aa}
\vspace{0.2cm} \hrule
\end{center}
\end{figure}

The ensuing values of Shannon and R\'{e}nyi entropies are depicted in
Fig.~{\ref{fig6aa}a} as functions of $\beta$.  Curve (i) is the Shannon ITUR, $\mathcal{H}(\mathcal{F}^{(2)}) + \mathcal{H}(\mathcal{F}^{(1)})$, and the dashed curve (ii) depicts the bound for the Shannon ITUR, i.e. $\log_2(e\pi)$. We see that the bound is saturated for small
$\beta$ (as should be expected for a single Gaussian wave packet) and gets worse as $\beta$ is increased  (and information about the localization worsens), but eventually saturates
at some value above the bound (when two Gaussian wave packets no longer overlap).
The plateau is a consequence of the mentioned fact that the RE is immune to piecewise rearrangements of the distributions. Namely, a PDF consisting of two well separated wave packets has the same RE irrespective of the mutual distance. This holds true also for the associated $\mathcal{F}^{(1)}$ PDF.    

The other curves are for the R\'{e}nyi ITURs: (iii) is
$\mathcal{I}_{1/2}({\mathcal{F}^{(1)}}) + \mathcal{I}_{\infty}({\mathcal{F}^{(2)}})$ where the qualitative behavior is similar as in the Shannon case. In this situation we see that the plateau forms earlier, which indicates that information about the peak part (i.e. $\mathcal{I}_{\infty}$) starts to saturate earlier than in the Shannon entropy case, which  democratically takes into account all parts of the underlying PDF.
The dashed curve (iv) is the other way round, i.e. $\mathcal{I}_{1/2}({\mathcal{F}^{(2)}}) + \mathcal{I}_{\infty}({\mathcal{F}^{(1)}})$.
The faint solid line overlapping with the dashed line (iv) is the
R\'{e}nyi entropy bound, $\log_2(2\pi)$.
We see that both configurations
saturate the bound for small $\beta$, but the dashed one saturates the
bound for all $\beta$. The saturation of the information bound can be attributed to the interplay between the degradation of information on tthe ail parts of $\mathcal{F}^{(2)}$ carried by $\mathcal{I}_{1/2}({\mathcal{F}^{(2)}})$ and the gain of information on the central part of ${\mathcal{F}^{(2)}}$ conveyed by  $\mathcal{I}_{\infty}({\mathcal{F}^{(1)}})$. Interestingly enough, the rate of change (in $\beta$) for both REs is identical but opposite in sign  thus yielding a $\beta$-independent ITUR.

For the same reasons as in the previous subsection the R-ITUR outperforms the S-ITUR.
In addition, we again note that while the variance $(\Delta x_{\pi/2})^2$ is finite for
arbitrary $\beta$,
\begin{eqnarray}
\langle CSS_{+} (\beta)|(\triangle x_{\pi/2})^2|CSS_{+} (\beta)\rangle \ = \ ({N}^{+}_{\beta})^2 \left[1 + e^{-2\beta^2}(1-4\beta^2) \right],
\end{eqnarray}
the variance of the conjugate quadrature
\begin{eqnarray}
\langle CSS_{+} (\beta)|(\triangle x_0)^2|CSS_{+} (\beta)\rangle\ = \ ({N}^{+}_{\beta})^2
\left[1 + e^{-2\beta^2} + 4\beta^2 \right],
\end{eqnarray}
can be arbitrary large subject to the value of $\beta$ (see Fig.~{\ref{fig6aa}}b). In this respect the
Schr\"{o}dinger--Robertson VUR again tends to be uninformative
for large values of $\beta$ (i.e., for a large wave-packet separation). Fig.~{\ref{fig6aa}}b
shows the product of the quadrature
variances (solid curve) along with the Robertson--Schr\"{o}dinger bound (dashed curve).
We see that the bound is saturated for small coherent state
amplitudes ($\beta$) but gets progressively worse as $\beta$ is
increased.

\section{Conclusions and Outlook\label{SEc6}}

In this paper we have generalized the information theoretic
uncertainty relations that have been previously developed in
Refs.~\cite{Hirschman1957, Bialynicky-Birula1975, Deutsch1983,
Maassen1988, Uffink1990} to include generalized information measures
of R\'{e}nyi and RE-based entropy powers. To put some flesh on
the bones we have applied these generalized ITURs
to a simple two-level quantum system (in the discrete-probability case) and to
quantum-mechanical systems with heavy-tailed distributions and Schr\"{o}dinger cat states
(in the continuous-probability case). An improvement of the R\'{e}nyi ITUR over both
the Robertson--Schr\"{o}dinger VUR and the Shannon ITUR was demonstrated in all the
aforementioned cases.

In connection with the discrete-probability ITUR we have also
highlighted a geometric interpretation by showing that
the lower bound on information content (or uncertainty) inherent in the ITUR
is higher, the smaller is the distance to singularity of
the transformation matrix connecting eigenstates of the two
involved observables.

%
%
%

The presented ITURs hold promise precisely because a large part
of the structure of quantum theory has an information theoretic underpinning
(see, e.g., Refs.~\cite{Nielsen2000,Brukner:99}).
In this connection it should be stressed that, ITURs in general should
play a central r\^{o}le, for instance, in quantum cryptography or in
the theory of quantum computers, particularly in connection with quantum error-correcting codes,
communication and algorithmic complexities.
In fact, information
measures such as R\'{e}nyi's entropy are used not because of intuitively pleasing aspects of
their definitions but because there exist various (classical and quantum) coding
theorems~\cite{Campbell1965,Jizba:12} which endow them with an operational (that is, experimentally verifiable) meaning.
While coding theorems do exist for Shannon, R\'{e}nyi or Holevo entropies,
there are (as yet) no such theorems for Tsallis,  Kaniadakis, Naudts and other currently popular entropies.
The information theoretic significance of such entropies is thus not obvious, though in the literature one can
find, for instance, a Tsallis entropy based version of the uncertainty relations~\cite{Wlodarczyk:08}.

Though our reasoning was done in the framework of the classical (non-quantum)
information theory, it is perhaps fair to mention that there exist various
generalizations of R\'{e}nyi entropies to the quantum setting. Most
prominent among theses are Petz's quasi-entropies~\cite{Petz:86} and
Renner's conditional min-, max-, and collision
entropy~\cite{Renner:05}. Nevertheless, the situation in the quantum
context is much less satisfactory in that these generalizations do
not have any operational underpinning and, in addition, they are
incompatible with each other in number of ways. For instance,
whereas the classical conditional min-entropy can be naturally
derived from the R\'{e}nyi divergence, this does not hold for their
quantum counterparts. At present there is no obvious generalization of
the R\'{e}nyi entropy power in the quantum framework and hence it is
not obvious in what sense one should interpret the prospective
ITUR. All these aforementioned issues are currently under active
investigation.

Let us finally make a few comments concerning the connection of the
entropy power with Fisher information. Fisher information was
originally employed by Stam~\cite{Stam:59} in his proof of the Shannon
entropy power inequality. Interestingly enough, one can use either
the entropy power inequality or the Cram\'{e}r--Rao inequality and
logarithmic Sobolev inequality to re-derive the usual
Robertson--Schr\"{o}dinger  VUR. While the generalized ITUR
presented here can be derived from the generalized entropy power
inequality (as both are basically appropriate restatements of
Young's theorem), the connection with Fisher information (or some of
its generalizations) is not yet known.
%
The corresponding extension of our approach in this direction would be worth pursuing
particularly in view of the natural manner in which RE is used both in inference
theory and ITUR formulation.

Last, but not least, the Riesz--Thorin and Beckner--Babebko inequalities that we have utilized in
Sections~\ref{Sec3} and \ref{SEc4}
belong to a set of inequalities commonly known as
${\mathcal{L}}^p$-interpolation theorems~\cite{Reed1975}. It would
be interesting to see whether one can sharpen our analysis from
Section~\ref{Sec3} by using the Marcinkiewicz interpolation
theorem~\cite{Reed1975}, which in a sense represents the deepest interpolation theorem.
In particular, the latter avoids entirely the Riesz convexity
theorem which was key in our proof. Work along these lines is
presently in progress.

\section*{Acknowledgments}

P.J. would like to gratefully acknowledge stimulating discussions
with H.~Kleinert, P.~Harremo\"{e}s and D.~Brody. This work was supported
by GA\v{C}R Grant No. P402/12/J077.

\appendix
\section{}

In this Appendix we introduce the (generalized) Young inequality and derive some
related inequalities. Since the actual proof of Young's inequality is rather involved
we provide here only its statement. The reader can find the proof
together with further details, e.g., in Ref.~\cite{Lieb1976}.

\begin{theorem} [Young's theorem]
Let $q,p,r > 0$ represent H\"{o}lder triple, i.e.,
\begin{eqnarray*}
\frac{1}{q} + \frac{1}{p} = 1 + \frac{1}{r}\, ,
\end{eqnarray*}
and let ${\mathcal{F}} \in \ell^{q}({\mathbb{R}^{D}})$ and
${\mathcal{G}} \in \ell^{p}({\mathbb{R}^{D}})$ are two non-negative
functions, then
\begin{eqnarray}
|\!|{\mathcal{F}}\ast {\mathcal{G}}|\!|_r \geq C^D
|\!|{\mathcal{F}}|\!|_q|\!|{\mathcal{G}}|\!|_p\, , \label{a.0a}
\end{eqnarray}
for $q,p,r \geq 1$ and
\begin{eqnarray}
|\!|{\mathcal{F}}\ast {\mathcal{G}}|\!|_r \leq C^D
|\!|{\mathcal{F}}|\!|_q|\!|{\mathcal{G}}|\!|_p\, , \label{a.0b}
\end{eqnarray}
for $q,p,r \leq 1$. The constant $C$ is
\begin{eqnarray*}
C = C_pC_q/C_r \;\;\;\;\;\; \mbox{with} \;\;\;\;\;\; C^2_x =
\frac{|x|^{1/x}}{|x'|^{1/x'}}\, .
\end{eqnarray*}
Here $x$ and $x'$ are H\"{o}lder conjugates. Symbol $\ast$ denotes a convolution.
\end{theorem}

Young inequality allows to prove very quickly the Hausdorff--Young
inequalities which are instrumental
in obtaining various Fourier-type uncertainty relations. In fact, the following
chain of reasons holds
\begin{eqnarray}
|\!| {\mathcal{F}}\ast\delta |\!|_r \geq C^D |\!|
{\mathcal{F}}|\!|_q |\!| \delta|\!|_p  = C^D |\!|
{\mathcal{F}}|\!|_q V_R^{(p-1)/p}\, . \label{a.1}
\end{eqnarray}
Here we have used the fact that for the $\delta$ function
\begin{eqnarray*}
|\!| \delta|\!|_p = \left[\int_{{\mathbb{R}}^D}  d{{\boldsymbol{x}}} \ \!
\delta^{p}({\boldsymbol{x}})\right]^{1/p} = \ \!\left[\int_{{\mathbb{R}}^D}
d{{\boldsymbol{x}}} \ \! \delta({\boldsymbol{x}})\delta^{p-1}(0)\right]^{1/p} = \ \!
V_R^{(p-1)/p}\, .
\end{eqnarray*}
In the derivation we have utilized that
\begin{eqnarray*}
\delta(0) = \int_{{\mathbb{R}}^D} d{{\boldsymbol{x}}} \ \! e^{i {{\boldsymbol{p}}} \cdot
{{\boldsymbol{0}}}} = V_R\, .
\end{eqnarray*}
Subindex $R$ indicates that the volume is regularized, i.e., we
approximate the actual volume of ${\mathbb{R}}^D$ with a
$D$-dimensional ball of the radius $R$, where $R$ is arbitrarily
large but fixed. At the end of calculations we send $R$ to infinity.
We should also stress that in (\ref{a.1}) an implicit assumption was
made that $q,p,r \geq 1$.

The norm $|\!| {\mathcal{F}}\ast\delta |\!|_r$ fulfills yet another
inequality, namely
\begin{eqnarray}
|\!| {\mathcal{F}}\ast\delta |\!|_r = \left[\int_{{\mathbb{R}}^D}
d{{\boldsymbol{x}}} \ \!\left( \int_{{\mathbb{R}}^D} d {{\boldsymbol{p}}} \ \! e^{- i {\bf
p} \cdot {\boldsymbol{x}}} \hat{\mathcal{F}}({\boldsymbol{p}}) \right)^{\! r \ \!}
\right]^{1/r}\leq |\!|\hat{\mathcal{F}}|\!|_n V_R^{1/n' + 1/r}\, ,
\label{a.2}
\end{eqnarray}
where we have used the H\"{o}lder inequality
\begin{eqnarray*}
\int_{{\mathbb{R}}^D} d {{\boldsymbol{p}}} \ \! e^{- i {{\boldsymbol{p}}} \cdot {\boldsymbol{x}}}
\hat{\mathcal{F}}({\boldsymbol{p}}) = \left|\int_{{\mathbb{R}}^D} d {\boldsymbol{p}} \
\! e^{- i {\bf p} \cdot {\boldsymbol{x}}} \hat{\mathcal{F}}({\boldsymbol{p}})  \right|
\ \! \leq \ \!|\!|\hat{\mathcal{F}}|\!|_n\ \!|\!|e^{- i {\boldsymbol{p}}
\cdot {\boldsymbol{x}}}|\!|_{n'} = |\!|\hat{\mathcal{F}}|\!|_n V_R^{1/n'}\, ,
\end{eqnarray*}
with $n$ and $n'$ being H\"{o}lder's conjugates ($n\geq1$).

Comparing (\ref{a.1}) with (\ref{a.2}) gives the inequality
\begin{eqnarray}
|\!|\hat{\mathcal{F}}|\!|_n V_R^{1/n' +1/r} \ \!\geq \ \!C^D |\!|
{\mathcal{F}}|\!|_q V_R^{(p-1)/p}\, . \label{a.3}
\end{eqnarray}
The volumes will mutually cancel provided $1/n' + 1/r + 1/p= 1$, or
equivalently, when $1/n' = 1/q -2/r$. With this we can rewrite
(\ref{a.3}) as
\begin{eqnarray}
|\!|\hat{\mathcal{F}}|\!|_n \ \! \geq \ \! C^D |\!|
{\mathcal{F}}|\!|_q \ \! \geq \ \! C^D |\!| {\mathcal{F}}|\!|_{n'}\,
. \label{a.4}
\end{eqnarray}
The last inequality results from H\"{o}lder's inequality:
\begin{eqnarray}
|\!| {\mathcal{F}}|\!|_a \ \! \geq \ \!|\!| {\mathcal{F}}|\!|_{b}
\;\;\;\;\;\; \mbox{when}\;\; \;\;\;\; a\leq b\, . \label{a.3b}
\end{eqnarray}
In fact, in the limit $r\rightarrow \infty$ the last inequality in
(\ref{a.4}) is saturated and $C\stackrel{r\rightarrow \infty
}{\rightarrow} 1$. Consequently we get the Hausdorff--Young inequality in the form
\begin{eqnarray}
|\!|\hat{\mathcal{F}}|\!|_n \ \!\geq  \ \! |\!|
{\mathcal{F}}|\!|_{n'}\, . \label{a.5}
\end{eqnarray}
This inequality holds, of course, only when $q\geq n'$ (cf. equation
(\ref{a.3b})), i.e., when $n \geq q/(q-1)$. Since $q\geq 1$ we have that $n \in [1,2]$.
Should we have started in our derivation with $\hat{\mathcal{F}}$ instead of
${\mathcal{F}}$ we would have obtain the reverse inequality
\begin{eqnarray}
|\!|{\mathcal{F}}|\!|_n \ \!\geq  \ \! |\!|
\hat{\mathcal{F}}|\!|_{n'}\, . \label{a.6}
\end{eqnarray}
Inequalities, (\ref{a.5}) and (\ref{a.6}) are known as classical
Hausdorff--Young inequalities~\cite{Hardy1959}. Note that in the spacial case when $n=2$ we have also
$n'=2$ and equations (\ref{a.5}) - (\ref{a.6}) together imply
equality:
\begin{eqnarray}
|\!|{\mathcal{F}}|\!|_2 \ \! =  \ \! |\!|
\hat{\mathcal{F}}|\!|_{2}\, . \label{a.7}
\end{eqnarray}
This is known as the Plancherel (or Riesz--Fischer)
equality~\cite{Hardy1959,Yosida1968}.

\begin{figure}
[ptb]
\begin{center}
\includegraphics[
height=2.3601in, width=2.6662in ] {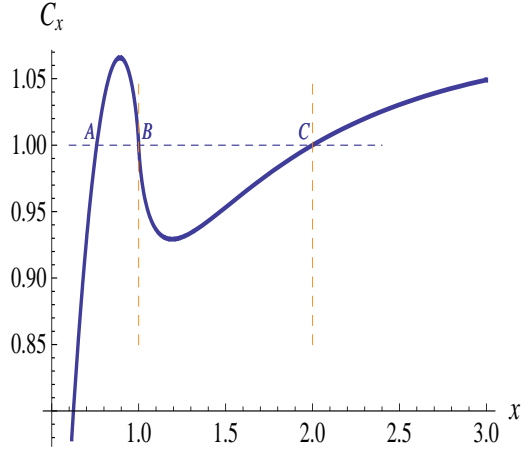} \caption{Dependence of the constant $C_x$ on the H\"{o}lder parameter
$x$. When $x$ is between points $B$ and $C$, i.e., when $x\in [1,2]$ than $C_x\leq 1$. For $x \leq A$
is $C_x$ also smaller than $1$ but such $x$ are excluded by the fact
that $x$ must be $\geq 1$.} \label{fig3}
\end{center}
\end{figure}

It should be noted that the Beckner--Babenko inequality from Section~5 improves upon the
Hausdorff--Young inequalities. This is because $C_x \leq 1$ for $x
\in [1,2]$, see Fig.~\ref{fig3}.
%
%
The Beckner--Babenko inequality follows
easily from Young's inequality. Indeed, assume that there exists a
(possibly $p$-dependent) constant $k(p) \leq 1$, such that
\begin{eqnarray}
k(p)|\!|{\mathcal{F}}|\!|_p \ \!\geq  \ \! |\!|
\hat{\mathcal{F}}|\!|_{p'}\;\;\;\;\;\; \mbox{and} \;\;\;\;\;\;
k(p)|\!|\hat{\mathcal{F}}|\!|_p \ \!\geq  \ \! |\!|
{\mathcal{F}}|\!|_{p'}\, . \label{a.8}
\end{eqnarray}
The constant $k(p)$ can be easily found by writing
\begin{eqnarray}
k(r)|\!| {\mathcal{F}}\ast {\mathcal{G}} |\!|_r \ \! \geq \ \! |\!|
\hat{\mathcal{F}}\hat{\mathcal{G}} |\!|_{r'} \ \! \geq \ \!|\!|
\hat{\mathcal{F}}|\!|_{q'}|\!|\hat{\mathcal{G}} |\!|_{p'} \ \! \geq
\ \!  [k(q')]^{-1}|\!|{\mathcal{F}}|\!|_{q} [k(p')]^{-1}
|\!|{\mathcal{G}}|\!|_{p}\, , \label{a.9}
\end{eqnarray}
which gives
\begin{eqnarray}
|\!| {\mathcal{F}}\ast {\mathcal{G}} |\!|_r \ \! \geq \ \!
[k(r)]^{-1}[k(q')]^{-1}[k(p')]^{-1}|\!|{\mathcal{F}}|\!|_{q}|\!|{\mathcal{G}}|\!|_{p}\,
. \label{a.10}
\end{eqnarray}
The middle inequality in (\ref{a.9}) is the H\"{o}lder inequality
that is valid for $1/r' = 1/p' + 1/q'$ (i.e., for $1/p + 1/q = 1/r +
1$). Comparison of (\ref{a.10}) with (\ref{a.0a}) gives the equation
\begin{eqnarray}
[k(r)]^{-1}[k(q')]^{-1}[k(p')]^{-1} = C^D = [C_qC_p/C_r]^D =
[1/C_{q'}C_{p'}C_r]^D\, .
\end{eqnarray}
This is clearly solved with $k(p) = C^D_p$. By choosing $p\in[1,2]$
we get improvement over the Hausdorff--Young inequalities.


\begin{thebibliography}{99}
\bibitem{C-L distribution} This distribution is in physics also known as the
Breit--Wigner distribution. It is particularly relevant in nuclear and
particle physics, and in spectroscopy.
\bibitem{footnote1} Note that, since
$a_{ij} \in SO(n)$, then $a_{ij}^{-1} = a_{ji}$ and thus
$\max_{ij}|a_{ij}^{-1}| =c$.
\end{thebibliography}
\end{document}